\documentclass[preprint]{emulateapj}
\usepackage{graphicx}
\begin{document}

\title{Constraining the Projected Radial Distribution of Galactic Satellites\\ with the Sloan Digital Sky Survey}
\author{Jacqueline Chen\altaffilmark{1}, Andrey V. Kravtsov\altaffilmark{1,2}, Francisco Prada\altaffilmark{3}, Erin S. Sheldon\altaffilmark{4}, Anatoly A. Klypin\altaffilmark{5},\\
 Michael R. Blanton\altaffilmark{4}, Jonathan Brinkmann\altaffilmark{6}, \& Aniruddha R. Thakar\altaffilmark{7} }

\altaffiltext{1}{Dept. of Astronomy and Astrophysics,
       Kavli Institute for Cosmological Physics,
       The University of Chicago, Chicago, IL 60637;
       {\tt jchen,andrey@oddjob.uchicago.edu}}
\altaffiltext{2}{Enrico Fermi Institute, The University of Chicago, Chicago, IL 60637}
\altaffiltext{3}{Ram\'on y Cajal Fellow, Instituto de Astrofisica de
  Andalucia (CSIC), E-18008 Granada, Spain; {\tt fprada@iaa.es}}
\altaffiltext{4}{Center for Cosmology \& Particle Physics, Dept. of Physics, New York University, 4 Washington Place, New York, NY 10003;  {\tt erin.sheldon,michael.blanton@gmail.com}}
\altaffiltext{5}{Astronomy Dept., New Mexico State University, MSC 4500, P.O. Box 30001, Las Cruces, NM 88003-8001;  {\tt aklypin@nmsu.edu}}
\altaffiltext{6}{Apache Point Observatory, P.O. Box 59, Sunspot, NM 88349;  {\tt jb@apo.nmsu.edu}}
\altaffiltext{7}{Department of Physics \& Astronomy, The Johns Hopkins University, 3400 North Charles Street, Baltimore, MD 21218;  {\tt thakar@jhu.edu}}

\begin{abstract}
  
  We use the Sloan Digital Sky Survey (SDSS) spectroscopic sample to
  constrain the projected radial distribution of satellites around
  isolated $\sim L_{\ast}$ galaxies.  We employ mock galaxy catalogs
  derived from high-resolution cosmological simulations to investigate
  the effects of interloper contamination and show that interlopers
  significantly bias the estimated slope of the projected radial
  distribution of satellites. We also show that the distribution of
  interlopers around galaxies is expected to be non-uniform in
  velocity space because galaxies are clustered and reside in crowded
  environments. Successful methods of interloper contamination
  correction should therefore take into account environments of the
  host galaxies.  Two such  new methods are presented and the most
  reliable of them is used to correct for interloper contamination in
  analyses of the SDSS galaxy sample. The best fit power-law slope of
  the interloper-corrected surface density distribution of satellites,
  $\Sigma(R)\propto R^{\alpha}$, in the volume-limited SDSS sample is
  $\alpha \simeq -1.7 \pm 0.1$, independent of the galaxy and
  satellite luminosities. Comparison with $\Lambda$CDM simulations
  shows that the radial distribution of the SDSS satellites is more
  concentrated than that of subhalos around galaxy-sized halos,
  especially at $R<100h^{-1}$~kpc. The predicted dark matter radial
  distribution is somewhat more concentrated than the profile of  the
  SDSS satellites, but the difference is not statistically significant
  for our sample.

\end{abstract}

%---------------------
\section{Introduction}
%---------------------
\label{sec:intro}

In the Cold Dark Matter (CDM) paradigm, satellite galaxies are
expected to be associated with the dark matter subhalos -- halos which
lie within the virial radius of a larger halo -- ubiquitous in the
cosmological CDM simulations. The abundance and radial distribution of
satellite galaxies can therefore serve as a useful test of CDM galaxy
formation models, constraining the relation between galaxies and subhalos.
In addition, satellite dynamics can provide useful constraints on the
total mass distribution in galactic halos
\citep[e.g.,][]{zaritsky_white94,zaritsky_etal97,prada_etal03,vandenbosch_etal04,conroy_etal05}.
This, however, requires a good understanding of how the spatial
distribution and kinematics of satellites and dark matter are related.

Many recent studies based on numerical simulations have shown that the
radial distribution of subhalos in cluster-sized systems is less
concentrated than that of dark matter in the inner $\approx 20-50\%$
of the virial radius of host halos, but approximately follows the dark
matter distribution at larger radii
\citep{ghigna_etal98,colin_etal99,ghigna_etal00,
  springel_etal01,delucia_etal04,diemand_etal04,gao_etal04,nagai_kravtsov05}.
Theoretical predictions for galaxy distributions in clusters have also
been accompanied by rapidly improving observational measurements
\citep[e.g.,][]{lin_etal04,hansen_etal05,collister_lahav05,yang_etal05,coil_etal05},
which also find concentrations of galaxy radial profiles lower than
the concentrations expected for the matter distribution of their
parent halos.

The observed distribution of satellite galaxies in galactic halos has
been studied less extensively.  The Local Group dwarf population is
more radially concentrated than subhalos in dissipationless numerical
simulations \citep{kravtsov_etal04b,taylor_etal04,willman_etal04}, a
bias that is likely related to the physics of the formation of the
smallest dwarf galaxies \citep{kravtsov_etal04b,diemand_etal05}. The
known population of the Local Group satellites is, however, quite
small compared to the expected population of CDM subhalos
\citep{klypin_etal99,moore_etal99}. Moreover, the strong radial bias
exhibited by the faint Milky Way satellites is not expected to apply
to the brighter satellites (such as, for example, the Magellanic
Clouds).

More accurate, statistical constraints on the satellite distribution can be
obtained by using galaxy redshift surveys.
Several early studies attempted to constrain the small-scale galaxy 
correlation function by estimating the surface density of objects 
projected near galaxies, $\Sigma(R) \propto R^{\alpha}$, finding slopes 
ranging from $\alpha=-0.5$ to $-1.25$ \citep{lake_tremaine80,phillipps_shanks87,vader_sandage91,lorrimer_etal94,smith_etal04,madore_etal04}.  

Recently, the availability of large galaxy redshift surveys has
allowed construction of large statistical samples of parent galaxies
and satellites with well defined selection criteria.  The large sample
sizes and redshift information make it possible to understand the
biases and completeness of the sample.  In addition, isolation
criteria for the primaries can be introduced in order to reduce the
interloper contamination and simplify the interpretation of results.
% Changed sentence below in reference to comment 10.
\citet{vandenbosch_etal05} use mock galaxy redshift samples derived
from large cosmological simulations to develop an iterative method of 
interloper rejection for the Two Degree Field Galaxy Redshift 
Survey (2dFGRS) and find that the data is generally consistent with the 
dark matter profile at large projected radii, but conclude that 
incompleteness of close pairs in the survey prevent strong constraints.
In an independent analysis,
\citet{sales_lambas05} account for the close-pair bias in the data by
estimating completeness with control samples of objects that are not
physically bound to the primaries.  They estimate the power-law slope of
the satellite distribution to be $\alpha = -0.96 \pm 0.03$ for projected
radii between 20 and 500 $h^{-1}$ kpc with a
significant dependence on morphological type of the parent galaxies
($\alpha\approx -1.1$ for the early type, and $\approx -0.7$ for the
late type galaxies). Note, however, that these values of $\alpha$ are obtained
without any correction for interlopers.

Given that the satellite distribution can be directly probed only in
projection, with only limited information about positions of likely
satellites in three dimensions, one has to worry about contamination
by interlopers, the objects that are not true satellites but are
simply close to the parent due to projection. Unfortunately, in
practice it is often tricky to estimate and correct for the interloper
contamination. This is especially difficult if the redshift
information is absent as was the case in the earliest studies of the
satellite distribution. However, even in studies in which redshift
information is available, the interlopers are often neglected
\citep[e.g.,][]{sales_lambas05}. Nevertheless, as we show in this
paper, the effect of interlopers must be corrected for in order to
obtain an unbiased measurement of the satellite projected radial
distribution.

The simplest assumption one can make is that the surface density of
interlopers is uniform. The interloper contamination can then be
estimated by sampling the environments around random points in the
field. This method thus presumes that the volume around a random point on
the sky and in redshift space contains a representative density of
interlopers.  However, bright galaxies are strongly correlated in
space and thus can be expected to be preferentially located in crowded
environments. One may suspect, then, that the random points method can
underestimate the interloper number density around real galaxies.
Therefore, more sophisticated methods, which sample interlopers in the
environments similar to those of the primary galaxies, need to be
developed.

In this study, we develop two new methods to estimate the contribution of
interlopers to the surface density of satellites, which take into
account the clustering of parent galaxies.  We use cosmological
simulations to test different methods of interloper subtraction and
present detailed discussion of their strengths and weaknesses.  We
show that interloper contamination can significantly bias measurements
of the projected radial distribution of satellite galaxies. Proper
interloper subtraction is, therefore, a must in studies of the radial
distribution of satellites. We use the Sloan Digital Sky Survey (SDSS)
spectroscopic sample to measure the projected radial distribution of
satellites around nearby bright galaxies, corrected for
interlopers. We compare the result to the predictions for the dark
matter and subhalo distribution in the $\Lambda$CDM cosmology.

The paper is organized as follows.  In \S~\ref{sec:testing} we discuss
the interloper contamination and different methods of interloper
subtraction, testing each of them using mock satellite catalogs
derived from cosmological simulations.  We then describe our SDSS
spectroscopic galaxy sample and the selection of primaries and
satellites in \S~\ref{sec:sdss}. In \S~\ref{sec:data} we derive the
interloper-corrected surface density profile of satellites in
volume-limited and flux-limited SDSS samples, and their subsamples, 
and compare results to the $\Lambda$CDM cosmological simulations.
We also discuss comparisons to simulations results. Our main results
and conclusions are summarized in section \S\ref{sec:conclusions}.
Throughout this paper, we assume flat $\Lambda$CDM cosmology with
$\Omega_{\rm m}=0.3$ and $h=0.7$.

\begin{table*}[t]
\begin{center}
\caption{Selection \& Isolation Criteria for Test Samples\label{tab:select_test}}
\begin{tabular}{lccc}
\tableline \\
\multicolumn{1}{l}{Parameters} &
\multicolumn{1}{c}{Test Sample 1 (TS1)} & 
\multicolumn{1}{c}{Test Sample 2 (TS2)} &
\multicolumn{1}{c}{Test Sample 3 (TS3)} \\
\\
\tableline
\\
& \multicolumn{1}{l}{}\\
Constraints on primaries& $V_{\rm max}$ = 100-150, ..., 300-350 km s$^{-1}$& ~~$M_{r}<$ -20~~ & ~~$M_{r}<$ -20~~\\

Satellite objects
& DM particles & DM particles & satellite galaxies\\

Isolation criteria: \\
~~~Size difference & $V_{\rm max} > 0.5 V_{\rm max}^{\rm pri}$ & $V_{\rm max} > 0.5 V_{\rm max}^{\rm pri}$ & $\Delta M_{r} < 2$ \\
~~~Minimum projected distance, $\Delta R (h^{-1}$ Mpc)& 0.5 & 0.5 & 0.5\\
~~~Minimum velocity separation, $\Delta V$ (km ${\rm s^{-1}}$)& 1000 & 1000 & 1000 \\

Satellite sample criteria: \\
~~~Magnitude difference from the primary & ----- & ----- & $\Delta M_{r} > 2$ \\
~~~Maximum projected distance, $\delta r (h^{-1}$ Mpc)& 0.6 & 0.6 & 0.6\\
~~~Maximum velocity separation, $\delta v$ (km ${\rm s^{-1}}$)& 500 & 500 & 500, 1000\\

Number of isolated primaries&380, 289, 236, 143, 89& 728 & 728  \\

Number in satellite sample&7475, 11165, 16614,15354, 14899 & 50608 & 343, 401\\

Limiting magnitude $M_{r}$& ------ & ------ & -18\\
\\
\tableline
\end{tabular}
\end{center}
\end{table*}

%----------------------------------------------------------------------
\section{Interloper Subtraction}
%----------------------------------------------------------------------
\label{sec:testing}

We use cosmological $N$-body simulations to construct mock samples of
host halos and possible satellites, with which we can examine the
effects of interloper contamination on the satellite distributions and
test different methods of interloper subtraction.  We start with two
$80h^{-1}$~Mpc high-resolution dark matter simulations of the
concordance $\Lambda$CDM cosmology: $\Omega_m$=0.3, $h$=0.7, and
$\sigma_8$=0.9. The two simulations differ in the random seed of their
initial conditions. We use outputs of the simulations at a redshift of
0.1, which is similar to the average redshift of objects in the SDSS
spectroscopic survey.  The simulations were performed with the Adaptive
Refinement Tree (ART) $N$-body code
\citep{kravtsov_etal97,kravtsov99}.  Details of the simulations can be
found in \citet{tasitsiomi_etal04}.  Halo identification was performed
using a variant of the Bound Density maxima halo finding algorithm
\citep{klypin_etal99b}.  Details of the algorithm and parameters used
in the halo finder can be found in \citet{kravtsov_etal04}.

In addition, we use a simple, observationally motivated scheme to
assign luminosities to the halos.  Details of this method
can be found in \citet{tasitsiomi_etal04}.  Halos are assigned
luminosities by matching the cumulative velocity function, $n(> V_{\rm
  max})$, where $V_{\rm max}$ is the maximum circular velocity, to the
SDSS luminosity function, $n(< M_{r})$ at $z=0.1$
\citep{blanton_etal03}. $M_{r}$ is the SDSS $r$-band absolute
magnitude defined as $M-5{\rm log}h$.  The magnitudes have been
K-corrected to $z$=0.0, using {\tt kcorrect v3.2}
\citep{blanton_etal03}.  Scatter is introduced in the relation between
$V_{\rm max}$ and $M_{r}$, assuming a standard deviation of 1.5
magnitudes for the $M_{r}$ distribution at fixed $V_{\rm max}$.  All
galaxies down to $M_{r}$ = -18 (corresponding to a mean minimum
$V_{\rm max}$ = 100 km s$^{-1}$) are included. 

In the remainder of the paper, our terminology is to refer to the
sample of hosts or primaries constructed using isolated DM halos
(i.e., objects which do not lie within a virial radius of a larger
object) selected by their maximum circular velocities as a sample of
``halos'' and the sample of isolated DM halos assigned $r$-band
luminosities as a sample of ``galaxies,'' respectively.  
We use $V_{\rm max}$ to quantify the size of halos because it is measured
more robustly and not subject to the same ambiguity as mass
definitions.  For ``galaxies'', we use magnitudes, $M_{r}$, to
quantify the size of galaxies, as an alternative to $V_{\rm max}$ to
account for possible effects of scatter between $V_{\rm max}$ and
$M_{r}$.  

% Changed below in reference to comments 1 and 2.
For possible satellites, subhalos assigned $r$-band luminosities 
are referred to as 'satellite galaxies' and are selected by their $M_{r}$ 
magnitude difference from their parent 'galaxies.'  
For isolated halos, satellites are selected from an arbitrary 
 fraction of dark matter particles around each host halo.  These two
choices of test satellite samples are expected to bracket the possible
range of radial profiles of the real satellites
\citep{diemand_etal04,nagai_kravtsov05}. Below we detail the definition
of samples in our analysis. 

%-----------------------------------------
\subsection{Primary and Satellite Samples}
%-----------------------------------------

\begin{figure}[h]
\epsscale{1.18}
\plotone{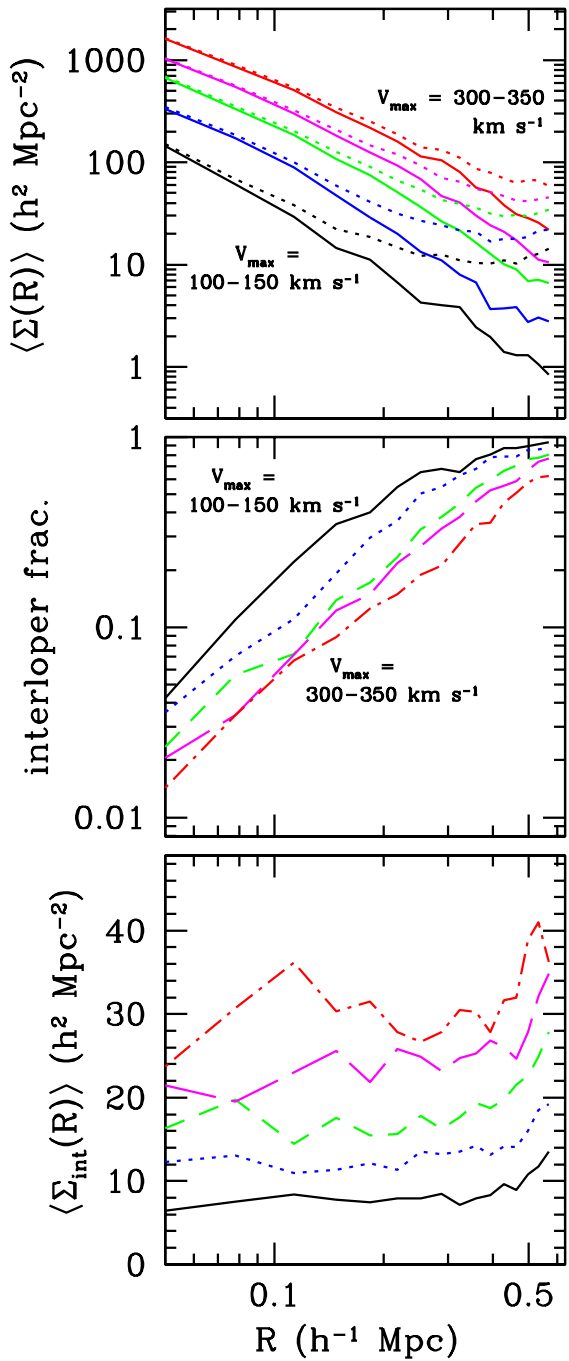}
\caption{Testing interloper subtraction with mock catalogs: 
  the derived average projected distribution of satellites compared to
  true satellites for the mock samples with five $V_{\rm max}^{\rm
    pri}$ ranges: 100-150, 150-200, 200-250, 250-300, and 300-350 km s$^{-1}$
  (Test Sample 1).  {\it Top:} The projected radial number density
  profile for true satellites (solid) and the satellite sample
  (dotted).  The amplitude in number density increases with $V_{\rm
    max}^{\rm pri}$.  {\it Center:} The fraction of interlopers in the
  satellite sample as a function of projected radius.  The bins are
  distinguished by line type: 100-150 (solid), 150-200 (dotted),
  200-250 (short dashes), 250-300 (long dashes), and 300-350
  (dot-dashed) km s$^{-1}$.  {\it Bottom:} The projected interloper number
  density profile using the same line types as the center panel.
  \label{fig:ints}}
\end{figure}

Previous studies of the projected distribution of galactic satellites
used observationally motivated selection criteria to construct a
primary sample of isolated host galaxies and a sample of
potential satellites that are projected close to primaries.  To test
several alternative methods for interloper subtraction, we construct a
set of primary and satellite samples derived from simulations using
different isolation criteria for the primaries and different selection
of the satellite samples.

For the primaries we use isolation criteria similar to those of
\citet{prada_etal03}.  We start by creating primary samples of halos
and galaxies.  An isolated primary halo, with circular velocity
$V_{\rm max}^{\rm pri}$, must have no other halos with a maximum
circular velocity $V_{\rm max} > 0.5 ~V_{\rm max}^{\rm pri}$ (which
corresponds to an absolute magnitude difference of $\approx 1.5-2$
magnitudes) within a projected separation $\Delta R=0.5 h^{-1}$~Mpc
and velocity separation, $\Delta V$ = $1000\ {\rm km\,s^{-1}}$. An
isolated galaxy primary with absolute magnitude $M_{r}^{\rm pri}$ must
have no other galaxies brighter than $M_{r}^{\rm pri}+2$, within the
same projected separation and velocity separation.

% Changed sentence below in reference to comment 2.
For each isolated primary, we construct two satellite samples. 
In the first, we use a random subset of DM particles found 
within a projected distance, $\delta r$ = 0.6 $h^{-1}$ Mpc,
and velocity difference, $\Delta v = 500\ {\rm km\,s^{-1}}$,
from each primary. These are our fiducial choices of 
$\delta r$ and $\delta v$. We also test a
non-fiducial velocity difference, $\delta v = 1000\ {\rm km\,s^{-1}}$.  
In the second satellite sample, we select all galaxies 
fainter than the primary by more than two magnitudes,
within the same projected distances and velocity
difference as above.

For our primary halos, we create several DM particle satellite samples
for different ranges of $V_{\rm max}^{\rm pri}$ (100-150, 150-200,
200-250, 250-300, 300-350 km s$^{-1}$).  This set of DM halo primaries and DM
particle satellites is referred to hereafter as Test Sample 1.  For
primary galaxies, we use a single sample of 728 galaxies with $M_{r} <
-20$ and build three different satellite samples using DM particles with the 
$\delta v$ = 500 km s$^{-1}$ criterion, satellite galaxies with $\delta v$ = 500 km s$^{-1}$, and satellite galaxies with $\delta v$
= 1000 km s$^{-1}$.  The set of primary galaxies and DM particle satellites
is referred to as Test Sample 2, while the set using satellite galaxies -- subhalos with assigned luminosities -- and both velocity criteria is labeled Test Sample 3.  These
samples are summarized in Table \ref{tab:select_test}.  For samples
with DM satellite particles, we bin the objects in radial bins of
35$h^{-1}$ kpc, starting at a minimum separation of 25.6$h^{-1}$ kpc.
For samples with satellite galaxies, we bin objects in bins of
70$h^{-1}$ kpc, starting at the same minimum separation. The larger
bin in the latter case is due to the smaller statistics of the subhalo
satellite sample.

%---------------------------------------------
\subsection{True Satellites vs. Interlopers}
%---------------------------------------------

There is a fraction of objects in our satellite samples that are not
gravitationally bound to the primaries but are included in the sample
because of projection effects. Throughout this paper, we call such
objects {\it interlopers}.    We define true
satellites as objects which satisfy the following negative binding
energy criterion:
\begin{equation}
E < \frac{1}{2}{\vert V\vert^2} + \phi(r), {\rm ~where~} \phi(r)= - \frac{v_{\rm esc}^2(r)}{2}.
\end{equation}
Here ${\bf V} = {\bf v}_{\rm sat} - {\bf v}_{\rm pri}$ is the 3D
satellite velocity with respect to the primary, and, for a
\citet[hereafter NFW]{navarro_etal97} profile,
\begin{equation}
v_{\rm esc}^{2} (r) = 9.26\, V_{\rm max}^{2} \frac{\ln (1+ r/r_{s})}{r/r_{s}},
\end{equation}
where $r_{s}$ has been estimated as $r_{s} = r_{\rm max}/2.15$ and $r_{\rm max}$ is the radius of the maximum circular velocity of the host halo, $V_{\rm max}$, as expected for a NFW density profile.

Using our test samples, we look for biases in the true satellite surface density profile and the projected distribution of the satellite sample.  We calculate the surface density of satellites per primary by
\begin{equation}
\langle \Sigma (R)\rangle = \frac{N_{\rm bin}}{\pi N_{\rm pri}(R_{2}^{2}-R_{1}^{2})},
\end{equation}
where $N_{\rm bin}$ is the number of objects in the satellite sample
that are found between the inner, $R_{1}$ and outer, $R_{2}$, radii of
the annulus and $N_{\rm pri}$ is the number of objects in the primary
sample.  $R$ is the midpoint of the bin.  

The projected number density profiles of all satellites and only true
satellites as a function of distance to the primary for Test Sample 1
are plotted in the top panel of Figure \ref{fig:ints}.  Here we can
see that interlopers significantly flatten the projected radial
distribution.  The flattening is the strongest for the surface
density profiles of satellite samples of the smallest primary halos.
For the different $V_{\rm max}^{\rm pri}$ primary samples, the DM
particle satellite samples show that the fraction of interlopers in
the satellite sample as a function of projected radius is similar for
all mass ranges, but decreases with increasing primary mass.  In
addition, at all masses, the projected number density of interlopers
(bottom panel of Fig. \ref{fig:ints}) is relatively flat, but rises at
radii greater than the isolation criterion.  The results for Test
Samples 2 and 3 are similar to those shown in Figure~\ref{fig:ints}.

%------------------------------------------
\subsection{Interloper Subtraction Methods}
%------------------------------------------

%------------------------------------
\subsubsection{Random Points Method}
%------------------------------------

\begin{figure}[t]
\epsscale{1.1}
\plotone{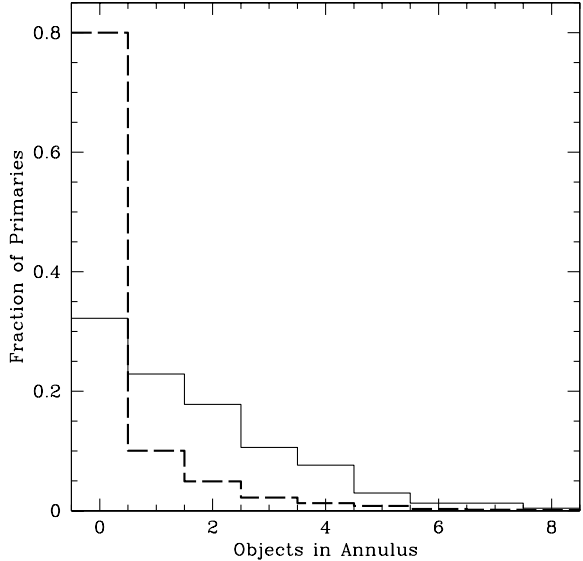}
\caption{The fraction of primaries with a number of objects with 
$V_{\rm max}>0.5V_{\rm max}^{\rm pri}$ in the
annulus of $0.5 < \Delta R <$1.0$h^{-1}$ Mpc and $\Delta V \leq 500\ 
{\rm km\,s^{-1}}$.  
The primary sample of $V_{\rm max}$ = 200-250 km s$^{-1}$ (Test Sample 1) is 
plotted here (thin, solid line) and the corresponding random points mock 
primary sample is shown in the thick, dashed line.  \label{fig:bkgnd}}
\end{figure}

\begin{figure}[t]
\epsscale{1.2}
\plotone{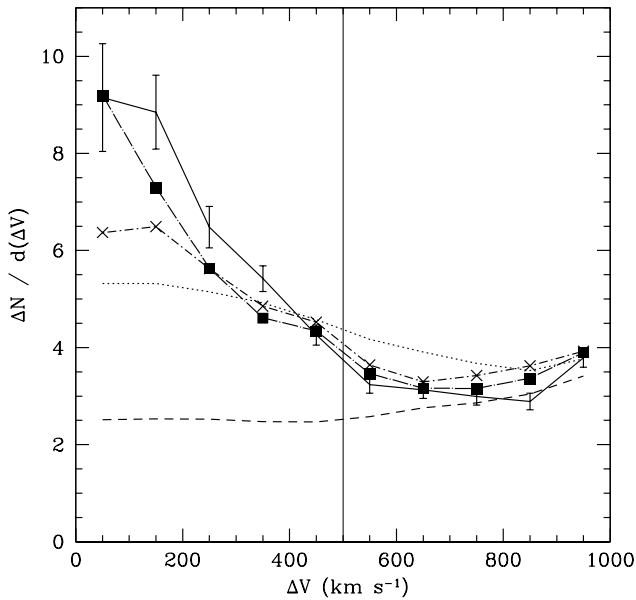}
\caption{Testing interloper substructure with mock catalogs: the relative velocity distribution of the interloper sample in the 
  mock catalog for primaries with $V_{\rm max}$ = 200 - 250 km s$^{-1}$ (Test
  Sample 1) for all DM particles within $25.6 < \Delta R< 500
  h^{-1}$ kpc (solid line, Poisson errors).  The estimated interloper
  sample for the clustered random points method is shown as a dotted
  line and for the random points method as a dashed line.  The nearby
  points methods are shown in dot-dashed lines, with square points for
  $\Delta R_{\rm corr}$ = 1 - 2$h^{-1}$ Mpc and crosses for $\Delta R_{\rm
    corr}$ = 1 - 5$h^{-1}$ Mpc. The thin vertical line shows the velocity
  difference criterion of the satellite sample of 500 km s$^{-1}$.
  \label{fig:gauss_int}}
\end{figure}

The most straightforward procedure for interloper removal, which we
call the random points method, assumes that the surface density of
interlopers is constant and can be estimated by surveying the area
around randomly placed points \citep[e.g.,][]{lake_tremaine80}. This
is one of the simplest methods to apply, since it can be implemented
even without redshift information.

We apply this method to our simulation-derived samples, by choosing
random points within the simulation box which satisfy the same
isolation criteria as the corresponding primary sample.  Each of these
random points, which make up a mock primary sample, has the
characteristics -- mass, magnitude, and velocity -- of an object in
the real sample of primaries.  This is accomplished by sampling the
primary sample with replacement -- i.e., assigning each random point
with the characteristics of an object in the primary sample, choosing
that object randomly from the entire primary sample.  To estimate the
interloper contribution, we construct a sample of random points that
has 20 times the number of objects in the sample of primaries. We can
expect, however, that uniformly distributed random points will
preferentially sample voids rather than the environments similar to
those of the isolated primaries. This is because 1) voids occupy most
of the volume and 2) our isolation criteria have forced the random
points to preferentially lie in voids, because the fraction of space
that satisfies our isolation criteria is larger in voids than in dense
environments.

This is illustrated in Figure~\ref{fig:bkgnd}, which shows the
distribution of massive ($V_{\rm max}>0.5V_{\rm max}^{\rm pri}$)
neighbors within the annulus described by the projected separation 
of $0.5 < \Delta R <$ 1.0$h^{-1}$ Mpc (i.e., just outside the radius
used in the isolation criteria) and 
velocity difference of $\Delta V = 500\ {\rm km\,s^{-1}}$
for our sample of primaries (Test Sample 1) and the sample of 
random points satisfying
the same isolation criteria. 
% Changed below in reference to Comment 3.
The Figure~\ref{fig:bkgnd} shows 
that primaries have systematically larger number of massive 
neighbors than the random points, and, therefore, that primaries occupy clustered regions of space.   The random points method
is thus expected to consistently underestimate
the interloper contamination.  This result may also be surmised from the 
bottom panel of Fig. \ref{fig:ints}, which shows that the number density of 
interlopers is not uniform and is correlated with primary halo mass.  
More massive isolated halos have more interlopers, which is 
consistent with the concept that, in general, more massive objects 
are more clustered.    

%---------------------------------------------
\subsubsection{Clustered Random Points Method}
%---------------------------------------------

It is clear that we can improve upon the random points method, if we
select not random points but points that are in environments similar
to those of primary galaxies.  We attempt to achieve this in the
clustered random points method.   Isolated points are chosen at random and the number of objects in
the annulus described above is calculated.  Points that allow the fractional
distribution of annular objects from the clustered random points to
exceed that of the primary points are rejected.  In other words, the method
insures that the mock sample contains the same distribution of massive neighbors around the primaries, as shown
in Figure~\ref{fig:bkgnd}.  As we discuss below,
the clustered random points method performs consistently better than
the random points method in estimating the interloper contamination.

 In Figure \ref{fig:gauss_int}, we show the relative velocity distribution of interlopers and estimated
interlopers for primaries with $V_{\rm max} = 200-250\ {\rm km\, s^{-1}}$ in Test Sample 1.  
  The interloper distribution is not constant as a function of velocity, showing a peak at $\Delta V = 0$ and a tail to $\Delta V$ = 1000 km $\rm s^{-1}$.  The interloper population
includes not just objects uniformly distributed in the velocity space,
but also objects with velocities correlated with
the velocity of the primaries. The estimated interloper population
using the random points is uniformly distributed in velocity space and
significantly underestimates the interloper contribution (dashed line). The
clustered random points method fares considerably better, although the
number of interlopers is still somewhat underestimated (dotted line).

%-----------------------------------
\subsubsection{Nearby Points Method}
%-----------------------------------

An alternative way of ensuring that the random points sample
environments of the primaries correctly is to pick isolated points
that are within a projected correlation length of real primaries.
\citet{smith_etal04} use such a method, estimating the background
using points at projected distances $>350 h^{-1}$ kpc.  In their
isolation criteria, they require that the magnitude difference between
a neighbor and the primary must be greater than 0.7 magnitudes for
galaxies within a projected distance of 700 $h^{-1}$ kpc.  The area
outside of a $350 h^{-1}$ kpc projected radius, however, may not have
the same isolation criteria as the primary galaxy sample.

Observational measurements of the two-point correlation function of
bright galaxies find a correlation length of $\approx 5h^{-1}$ Mpc
\citep[e.g.,][]{zehavi_etal04}.  In Figure~\ref{fig:gauss_int}, we
show the estimated fraction of interlopers using a method in which
test points are selected from the annuli of $\Delta R_{\rm corr} =1-2
h^{-1}$~Mpc and $\Delta R_{\rm corr} =1-5 h^{-1}$ Mpc around the
primaries.  We choose the inner radius of $1h^{-1}$~Mpc to avoid
sampling real satellites. All test points satisfy the same isolation
criteria as our sample of primaries.  The Figure~\ref{fig:gauss_int}
shows that the velocity distribution of interlopers for $\Delta R_{\rm
  corr} = 1-5 h^{-1}$ Mpc is somewhat similar to that of the clustered
random points method, while $\Delta R_{\rm corr} = 1-2 h^{-1}$ Mpc
choice recovers the true interloper velocity distribution much better.
We will, therefore, use this latter radial annulus as our fiducial
choice. As we show below, the nearby points method with the fiducial
$\Delta R_{\rm corr}$ is the best among the other methods we tested
here in recovering the interloper contamination in our mock samples.

%----------------------------------------
\subsubsection{"Gaussian $+$ Constant" Method}
%----------------------------------------

\begin{figure}[h]
\epsscale{1}
\plotone{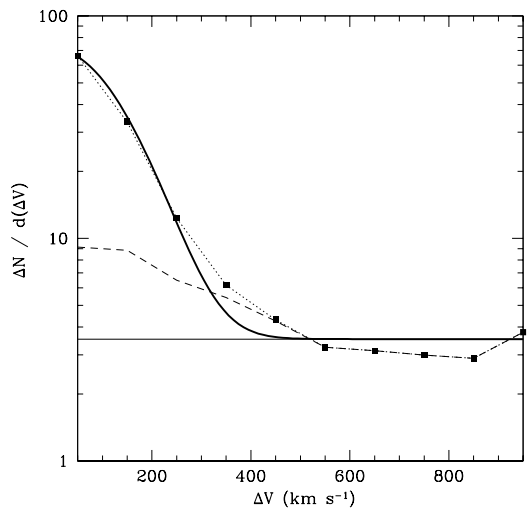}
%/home/jchen/satgals/Results/new/test_part_gal/mle/gauss_test.ps
\caption{The Gaussian plus constant method (thick, solid line) is compared to the velocity 
  distribution of the satellite sample (square points, dotted
  line) and the interloper sample (dashed line) for primaries with
  $V_{\rm max}$ = 200-250 km s$^{-1}$ and satellites with $25.6 < \Delta R <
  500 h^{-1}$ kpc (Test Sample 1).  The thin horizontal line shows the
  estimated interloper fraction from the Gaussian plus constant
  method.  \label{fig:gauss}}
\end{figure}

For completeness, we also test the interloper subtraction method used
by \citet{mckay_etal02} and \citet{prada_etal03}. The method does not
use random points, but assumes instead that 
the velocity distribution of satellites
can be described by a Gaussian, while the distribution of interlopers
is uniform. The velocity distribution of interlopers shown in
Figure~\ref{fig:gauss_int} is inconsistent with this assumption; a result 
that is in agreement with conclusions by \citet[][see their Fig.~1]{vandenbosch_etal04}. 

Nevertheless, the tests performed on our mock samples show that the
method does estimate the velocity dispersion of the true satellites --
the purpose for which the method was originally used by
\citet{mckay_etal02} and \citet{prada_etal03} -- quite accurately.
This is because the velocity dispersion of the interlopers, correlated
with primaries in velocity space, is similar to that of the
satellites.  Their inclusion into the satellite samples thus does not
bias the velocity dispersion appreciably.  
% Change below (added sentence) in ref. to Comment 4.
The velocity dispersion of interlopers is likely due to the infall of 
objects along filaments.  

Nevertheless, to correctly account for the interloper contribution to the
radial surface density profile this is not sufficient.  The assumption
that all of the objects correlated with the primary in velocity space
are true satellites will lead to an underestimate of interloper
contribution.  This is demonstrated in Figure~\ref{fig:gauss}. Given
that the interloper fraction increases with increasing projected
radius, the "Gaussian plus constant" method leads to surface density
profiles flatter than the true distribution.  Another problem occurs
when the number of satellites is small and their velocity distribution
is not well sampled.  We find that in such cases the
Gaussian$+$constant fits can be unstable.

In our tests, we use objects within $\delta v=1000\ {\rm km\,s^{-1}}$ to fit 
a gaussian$+$constant. This is larger than $\delta v=500\ {\rm km\,s^{-1}}$ used by \citet{prada_etal03} and in this paper fiducially, because our sample includes massive 
objects with velocity dispersion close to the $\delta v$, in which case 
the interloper constant is poorly constrained. 

%------------------------------------------------------ 
\subsection{Testing the Interloper Subtraction Methods}
%------------------------------------------------------

\begin{figure}[t]
\epsscale{1.1}
\plotone{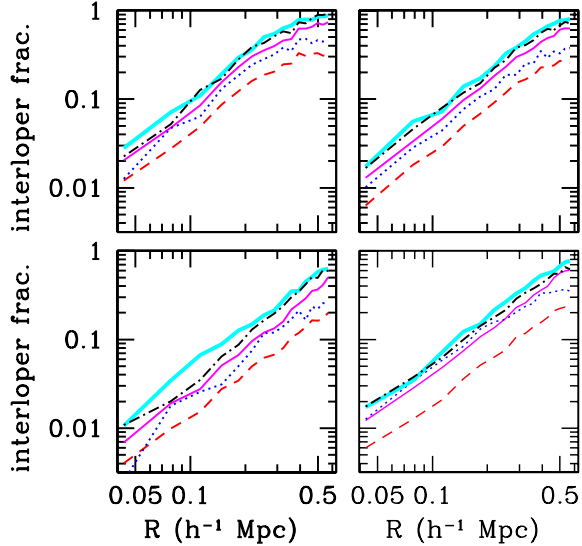}
\caption{Testing interloper subtraction with mock catalogs: the 
  fraction of interlopers for different mock satellite samples
  estimated by different methods.  Clockwise from the top, left panel,
  the $V_{\rm max}$ ranges are 150-200, 200-250, 250-300, and 300-350
  km s$^{-1}$, using DM particle satellites (Test Sample 1).  The
  corresponding number density profiles are shown in Fig.
  \ref{fig:ints} (top panel).  In all plots, the thick lines represent the true
  interloper fraction (also shown in Fig.
  \ref{fig:ints}, center panel), the dotted lines use the Gaussian plus constant
  method, the dashed lines use the random points method, the solid
  lines use the clustered random points method, and the dot-dashed
  lines use the nearby points method, with $\Delta R_{\rm
    corr}=1-2h^{-1}$~Mpc.
  \label{fig:int_sub}}
\end{figure}

We compare how well the four different interloper subtraction methods
described above recover the true projected density profile of
satellites $\langle\Sigma (R)\rangle_{\rm true}$, where
\begin{equation}
\langle\Sigma (R)\rangle_{\rm int ~sub} = 
\langle\Sigma (R)\rangle_{\rm sat}-\langle\Sigma (R)\rangle_{\rm est ~int},
\end{equation}
where our interloper subtracted density profile, $\langle\Sigma (R)\rangle_{\rm int ~sub}$, is the profile of the satellite sample, $\langle\Sigma (R)\rangle_{\rm sat}$, minus the interloper profile, $\langle\Sigma (R)\rangle_{\rm est ~int}$, estimated using different methods described above.   

Figure~\ref{fig:int_sub} shows the results for Test Sample 1, for
different $V_{\rm max}^{\rm pri}$ ranges of our primaries. The figure
shows that the random points method performs the worst.  The "Gaussian
plus constant" and clustered random points methods perform better,
although the latter does a better job overall in estimating the
interloper contamination. The nearby points performs best.  For Test
Sample 2 and 3, the results are similar, except for the "Gaussian plus
constant" method, which runs into difficulties with fitting correct
interloper level for Test Sample 3, in which the number of
satellites is small and the velocity distribution is under-sampled.
This indicates that the method should be applied only to the samples
in which the velocity distribution of satellites and interlopers is
sampled well.

%---------------------------------------
\subsection{Fits to the radial profiles}
%---------------------------------------

We now test how well the fits to the interloper subtracted radial
surface density profiles recover the profile of the true satellites.
The data is binned with bins of 35$h^{-1}$ kpc starting at a projected
radius of 25.6 $h^{-1}$ kpc for the DM particles as satellites, and
with bins of 70$h^{-1}$ kpc for subhalos with assigned luminosities
starting at a projected radius of 25.6$h^{-1}$ kpc.  We fit the number density
profile with a power law,
\begin{equation}
\Sigma (R) = A R^{\alpha},
\end{equation}
for bins with mean projected radius smaller than 0.5$h^{-1}$ Mpc.  We
find the best fit values of the slope, $\alpha$, and its errors
marginalized over the normalization of the power law, for the true
satellite sample, the satellite sample (true satellites $+$ interlopers), and for the
interloper subtracted profile obtained with each of the methods
discussed above. We measure the $1\sigma$ errors in slope, which
reflects the statistical Poisson errors of the density profiles. 

In general, the true DM particle satellite distribution should be
well-approximated by a projected NFW profile.  However, we do not
attempt to fit an NFW profile because, as we can see below, the
current observational samples cannot discriminate between the NFW and
a simpler power-law profile.  We choose our sample of primaries (Test Samples 2 \& 3) using
luminosities instead of $V_{\rm max}$ and stack the satellites of many
primaries, possibly mixing NFW profiles with different concentrations
and virial radii together.  In addition, due to the limited satellite
statistics and the fiber collisions in the SDSS we use a minimum
projected radius, similar to that used in the SDSS spectroscopic
sample (see below), which is of order of 10\% of the virial radius of
the primary galaxies. The minimum radius, then, may be larger than the
scale radius of the NFW profile.

As can be seen from Tables~\ref{tab:slope_dm}
and~\ref{tab:slope_subhalos}, the DM particle satellite sample (Test
Sample 2) and the galaxy satellite sample (Test Sample 3) have
different radial distributions.  Most of the DM particles in the
satellite sample are from the smooth distribution of the parent halo,
not subhalos, which explains the steep, $\alpha = -1.81 \pm 0.01$,
slope found for the DM particle satellite population. This is
consistent with the steep slope of $\approx -3$ (or $\sim -2$ in
projection) predicted for the 3-D density profile of the CDM halos at
large radii. The radial distribution of the satellite sample which uses
subhalos with assigned luminosities is considerably flatter $\alpha =
-1.34 \pm 0.12$.  At small projected radii, the profile is flattened
by tidal disruption of subhalos.

\begin{table}[h]
\begin{center}
\caption{Estimated Power-Law Slope for Sample with DM Particles (Test Sample 2)\label{tab:slope_dm}}
\begin{tabular}{lc}
\tableline\tableline
\\
\multicolumn{1}{l}{Input Data} &
\multicolumn{1}{c}{25.6 $< R < 500 h^{-1}$ kpc} \\
\\
\tableline
\\
~~~true satellites         &  $-1.815 \pm 0.007$ \\
~~~satellite sample        &  $-1.352 \pm 0.007$  \\
~~~random points           &  $-1.481 \pm 0.008$ \\
~~~clustered random points   &  $-1.570 \pm 0.008$ \\
~~~nearby points ($\Delta R_{\rm corr}$ = 1 - 2)&  $-1.686 \pm 0.009$ \\
~~~gaussian $+$ constant   &  $-1.533 \pm 0.007$ \\
\\
\tableline
\end{tabular}
\end{center}
\end{table}

\begin{table}[h]
\begin{center}
\caption{Estimated Power-Law Slope for Sample with Satellite Galaxies (Test Sample 3)\label{tab:slope_subhalos}}
\begin{tabular}{lc}
\tableline\tableline
\\
\multicolumn{1}{l}{Input Data} &
\multicolumn{1}{c}{25.6 $< R < 500 h^{-1}$ kpc } \\
\\
\tableline
\\
$\delta v$ = 500 km s$^{-1}$   & \\
~~~true satellites           &  $-1.34 \pm 0.12$ \\
~~~satellite sample          &  $-0.84 \pm 0.11$ \\
~~~random points             &  $-0.95 \pm 0.12$  \\
~~~clustered random points   & $-1.08 \pm 0.12$  \\
~~~nearby points ($\Delta R_{\rm corr}$ = 1 - 2)             & $-1.20 \pm 0.14$  \\
~~~gaussian $+$ constant      & $-0.95 \pm 0.11$  \\
$\delta v$ = 1000 km s$^{-1}$      & \\
~~~true satellites           &  $-1.34 \pm 0.12$ \\
~~~satellite sample          &  $-0.75 \pm 0.11$ \\
~~~random points   & $-0.95 \pm 0.12$  \\
~~~clustered random points   & $-1.23 \pm 0.14$  \\
~~~nearby points ($\Delta R_{\rm corr}$ = 1 - 2)    & $-1.24 \pm 0.15$  \\
~~~gaussian $+$ constant      & $-0.94 \pm 0.11$  \\
\\
\tableline
\end{tabular}
\end{center}
\end{table}

In Tables \ref{tab:slope_dm} and~\ref{tab:slope_subhalos}, we show
that the best-fit slopes for the satellite radial profile are significantly
flattened by interlopers.  Without any interloper subtraction, the
estimated surface density profile will be shallower than the profile
of the true satellite population, in both the DM particle sample and
the galaxy sample by $\approx 0.5$ in the power-law slope.

The random points method is inadequate for recovering the correct
slope: the recovered slope slopes are shallower than those of the true
distribution by $\approx 0.4$.  The "Gaussian plus constant" method 
also underestimates the slope significantly.
Interloper subtraction by the clustered random points and nearby
points perform reasonably well in all test samples.  In general, the
nearby points method performs the best giving on average a steeper
slope (by $\approx 0.1$) than the clustered random points method.
Overall, the interloper bias still persists as the slope
estimated with the nearby points method systematically 
underestimates the true slope by $\approx 0.1$, even though 
the slope values estimated with this method are 
within one standard deviation of the true slope for Test Sample 3. We will 
therefore use the nearby points method as our interloper correction
method of choice, keeping in mind that the best fit slope
should be corrected by $\Delta\alpha_{\rm bias}\approx 0.1$. 

%---------------------------------------
\section{The SDSS Spectroscopic Survey}
\label{sec:sdss}
%---------------------------------------

The Sloan Digital Sky Survey (SDSS) \citep{york_etal00} will image up
to $10^4$ deg$^2$ of the northern Galactic cap in five bands,
$u,g,r,i,z$, down to $r \sim 22.5$ 
\citep{fukugita_etal96,hogg_etal01,smith_etal02} using a
dedicated 2.5m telescope at Apache Point Observatory in New Mexico  \citep{gunn_etal98,gunn_etal06}.
In addition to the imaging survey, the SDSS main galaxy sample is a
subsample of objects from the imaging catalog which have been targeted
for spectroscopic observations \citep{strauss_etal02}.  The
spectroscopic targets are selected with $r$-band Petrosian magnitudes
$r \leq 17.77$ and $r$-band Petrosian half-light surface brightnesses
$\mu_{50} \leq 24.5$ mag arcsec$^{-2}$. The median redshift of the
SDSS main galaxy sample is 0.104.

The SDSS spectroscopy is carried out using 640 optical fibers
positioned in pre-drilled holes on a circular plate in the focal plane
of 3 degrees diameter, with minimum separation between fibers of
55$\arcsec$. Targeted imaging regions are assigned spectroscopic
plates by an adaptive tiling algorithm \citep{blanton_etal03b}, which
also assigns each object a fiber. An automated pipeline measures the
redshifts and classifies the reduced spectra \citep[D. J. Schlegel et
al. 2005, in preparation]{stoughton_etal02,pier_etal03,ivezic_etal04}.

For this catalog we use the reductions of the SDSS spectroscopic data
performed by D. J. Schlegel et al. (2005, in preparation) using their
reduction code, which extracts the spectra and finds the redshifts.
The redshifts found are, over 99\% of the time for Main galaxy sample
targets, identical to the redshifts found by an alternative pipeline
used for the SDSS Archive Servers (M. SubbaRao et al. 2005, in
preparation).

%Changed sentence below, not in response to the ref report, but 
%to avoid confusion.
For this analysis, we use a subset of the available spectroscopic main
galaxy sample released as of Data Release Four \citep{adelman_mccarthy_etal06}, but including all of the galaxy sample released as of Data Release Three \citep{abazajian_etal05}.  This catalog, known as LSS SAMPLE14, is built from the New York University Value-Added Galaxy Catalog \citep{blanton_etal05} and contains 312,777 galaxies.  
Because the SDSS spectroscopy is taken through
circular plates with a finite number of fibers of finite angular size,
the spectroscopic completeness varies across the survey area. The
resulting spectroscopic mask is represented by a combination of disks
and spherical polygons \citep{tegmark_etal04}.  Each polygon also
contains the completeness, a number between 0 and 1 based on the
fraction of targeted galaxies in that region which were observed. We
apply this mask to the spectroscopy and include only galaxies from
regions where the completeness is greater than 90\%, an area of 3448 square degrees. The same
criterion is applied for catalogs of clustered random points and
nearby points used for the interloper subtraction.

%-----------------------------------------------------------------

\section{Radial Distribution of Satellites Around SDSS Primaries}
\label{sec:data}
%-----------------------------------------------------------------

Unlike the numerical simulations where we have good resolution and
100\% completeness, when using spectroscopic data to stack objects and
estimate a surface density, we suffer from two major problems: fiber
collisions and incompleteness.

As described previously, the minimum separation between fibers, the
fiber collision separation, is 55$\arcsec$.  At a redshift of $z$ =
0.035, the approximate median of our SDSS sample of primary galaxies,
the fiber collision separation is 26.8 $h^{-1}$ kpc.  Fiber collisions
could bias the small projected radii end of the radial distribution,
removing objects that should be counted and tilting it shallower.
However, some of the area of the survey has been re-observed and the
overlap region could have objects observed with separations as small
as the fiber diameter, 3$\arcsec$.  In our samples, we do not use
objects at projected radii smaller than the fiber collision
separation.  

We use the $r$-band magnitudes in the LSS SAMPLE14 subsample of the
SDSS spectroscopic survey, normalized to $h$=1, such that $M_{r} =
M_{0.1_{r}} - 5{\rm log}_{10} h$, where $M_{0.1_{r}}$ is the absolute
magnitude K-corrected to $z$=0.1 as described in
\citet{blanton_etal03c}.  The LSS SAMPLE14 also provides measures of
the fraction of objects with spectra in the area of an object.  While
fiber collisions remove objects from our survey, incompleteness
removes area from our survey, which would likely tilt the number
density steeper, since there is more area further away from a primary
galaxy.  Constraining our sample to include only galaxies with a
minimum completeness fraction of 90\% should lead us to be complete to
that level.  However, since we need to search the area around each
primary galaxy, it is possible that the search area will not be
contained on an area of one completeness level but can overlap with an
area of a lower completeness percentage.  A simple check of this is to
calculate an analogue to the projected cross-correlation function,
$w(R)$, which is unbiased by incompleteness and can be compared to the
projected number density estimated using the random points method.  
We apply such test and show that incompleteness does not bias
our estimates of the radial profiles (see \S~\ref{sec:completeness}). 

%Change here (added small paragraph) in reference to comment 6. 
We create both a volume-limited sample (\S~\ref{sec:volume}) and a flux-limited sample (\S~\ref{sec:flux}).  The flux-limited sample offers 
better statistics, but it is also biased to brighter satellite objects and the 
radial profile of satellite galaxies has not been established to be 
independent of the magnitude of satellites.  For this reason, we also 
test a volume-limited sample with poorer statistics.

%-----------------------------------
\subsection{Volume-limited samples}
\label{sec:volume}
%-----------------------------------

\begin{table*}[t]
\begin{center}
\caption{Selection \& Isolation Criteria for SDSS Samples\label{tab:select_sdss}}
\begin{tabular}{lcc}
\tableline \\
\multicolumn{1}{l}{Parameters} &
\multicolumn{1}{c}{~~~Volume-Limited~~~} & 
\multicolumn{1}{c}{~~~Flux-Limited~~~} \\
\\
\tableline
\\
Maximum depth of sample (km ${\rm s^{-1}}$)....................& 13,500 & 13,500 \\
Constraints on primaries.......................................& -23 $<M_{r}<$ -20 & -23 $<M_{r}<$ -20\\
Constraints on bright neighbors: \\
~~~Maximum magnitude difference........................& 2& 2\\
~~~Minimum projected distance, $\Delta R (h^{-1}$ Mpc)....& 0.5 & 0.5 \\
~~~Minimum velocity separation, $\Delta V$ (km s$^{-1}$)........& 1000 & 1000 \\

Constraints on satellites: \\
~~~Minimum magnitude difference.........................& 2& 2\\
~~~Maximum projected distance, $\delta r (h^{-1}$ Mpc)......& 0.5 & 0.5 \\
~~~Maximum velocity separation, $\delta v$ (km ${\rm s^{-1}}$)......& 500, 1000 & 500, 1000\\
~~~Minimum projected distance, $\delta r (h^{-1}$ Mpc)......& 0.329, 0.341 & 0.329, 0.341 \\

Number of isolated primaries................................&871 & 871\\

Number in satellite sample....................................&336, 357& 678, 786\\

Limiting magnitude $M_{r}$........................................& -17.77, -17.85& ----- \\
\\
\tableline
\end{tabular}
\end{center}
\end{table*}

\begin{figure}[h]
\epsscale{1.1}
\plotone{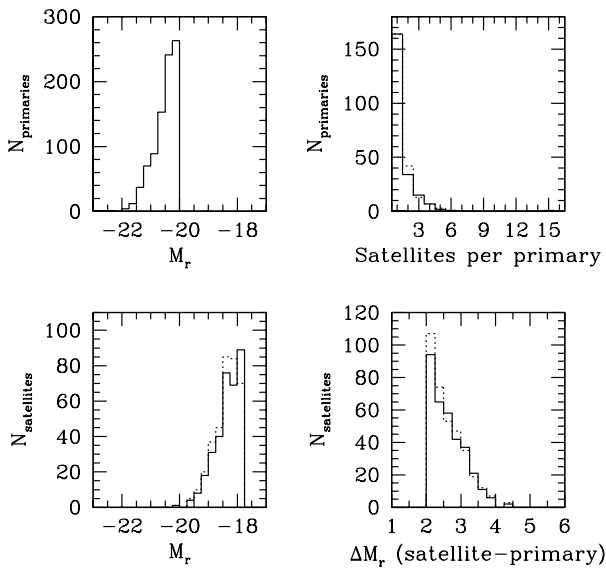}
%/home/jchen/satgals/data/erin_sdsscat/DAT/hist.ps
\caption{Statistics of primaries and satellites for a volume-limited sample.  {\it Top-left:}  The $r$-band magnitude histogram for primaries.  {\it Bottom-left:}  The $r$-band magnitude histogram for satellites.  The solid line shows the results for the $\delta v$ = 500 km s$^{-1}$ criterion, while the dotted line shows the $\delta v$ = 1000 km s$^{-1}$ criterion.  {\it Bottom-right:}  The magnitude differences between satellites and primaries, with line styles as in the bottom-left panel.  {\it Top-right:}  The number of satellites per primary for primaries with at least one satellite, with line styles as in the bottom-left panel.  \label{fig:hist}}
\end{figure}

\begin{figure}[h]
\epsscale{1.2}
\plotone{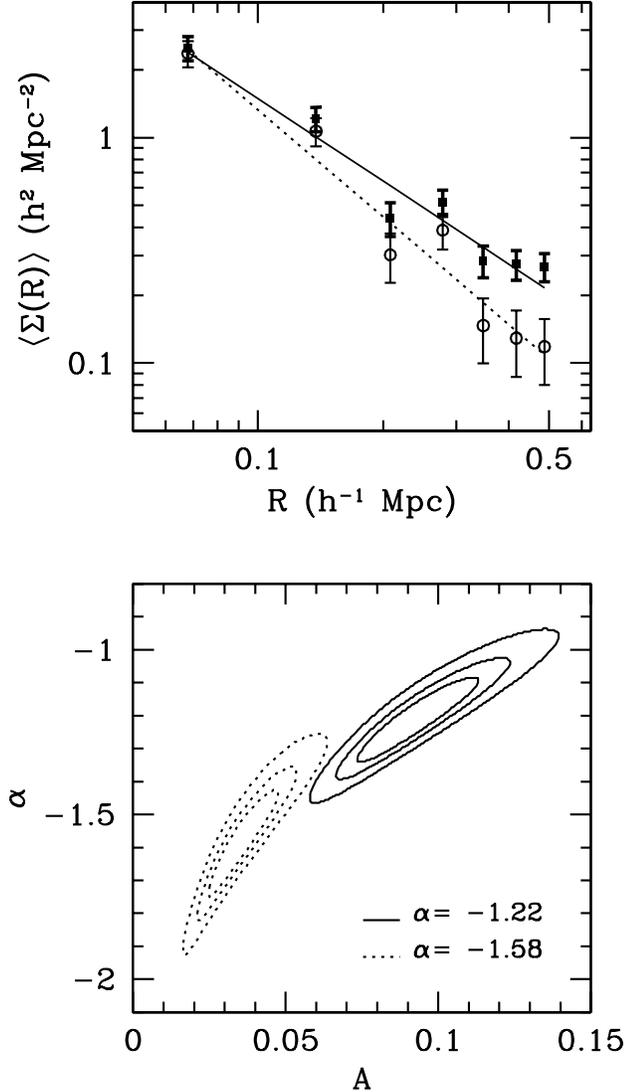}
%/home/jchen/satgals/data/erin_sdsscat/DAT/fit_data.ps
\caption{The satellite sample compared to interloper subtracted samples for the volume-limited sample using $\delta v$ = 500 km s$^{-1}$.  {\it Top:}  The best-fit power-law for the satellite sample (solid) and the interloper subtracted profile for the $\Delta R_{\rm corr}$ = 1 - 2 $h^{-1}$ Mpc nearby points (dotted).  The data is shown with error bars: satellite sample (square points, thick Poisson error bars) and nearby points (open circles, thin error bars).   {\it Bottom:} The 68\%, 90\%, and 99\% confidence intervals for the satellite sample (solid lines) and the $\Delta R_{\rm corr}$ = 1 - 2 $h^{-1}$ Mpc nearby points sample (dotted lines).  
%Change here;  comment 7.
The contours are labeled by the unmarginalized best-fit slope of each profile. \label{fig:fit_data}}
\end{figure}

We first create a volume-limited galaxy sample from the LSS SAMPLE14
with a depth of 13,500 km s$^{-1}$, corresponding to the limiting redshift of 
$z=0.045$. This limit is chosen as a trade-off between the volume of the 
sample and the absolute magnitude limit for our satellites, which 
would need to be decreased to brighter magnitudes for more distant 
primaries. The trade-off is also with the minimum separation at which fiber-collisions
become important, which increases with distance. To include more 
distant primaries we would have to sacrifice the ability to probe 
density distributions at small separations.

In total we have 21,851 galaxies. Since the isolation criterion
requires that we search for objects that are within $\Delta V$ = 1000
km s$^{-1}$, we can only search for primaries within the subset of velocities
1000 to 12,500 km s$^{-1}$.  For the satellite catalog to be volume-limited,
this requires a maximum absolute magnitude of $M_{\rm r,lim} - 5 {\rm
  log} h = 17.77 - DM -K_{0.1}$ in the $r$-band, where the 17.77 is
the flux limit in this band, $DM$ is the distance modulus, and
$K_{0.1}$ is the K-correction at $z$=0.1.  We use the $K$-correction
at $z=0.1$ in order to avoid underestimating the limiting absolute
magnitude.  As in the simulations, we test both the $\delta v$ = 500
km s$^{-1}$ satellite criterion and a larger $\delta v$ = 1000 km s$^{-1}$.  For
those limits, the minimum separation between fibers is 32.9 $h^{-1}$
kpc and 34.1 $h^{-1}$ kpc, respectively and the limiting absolute
magnitudes are -17.77 and -17.85.  The satellites are thus 
limited to the brightest satellite
galaxies, $\sim 0.1L_{*}$.  We choose galaxies that are in areas that
are at least 90\% complete and set the size of the mock primary sample
to be 20 times the number of primary galaxies.

\begin{table}[h]
\begin{center}
\caption{Estimated Power-Law Slopes for the Volume-Limited Sample\label{tab:slope_all}}
\begin{tabular}{lc}
\tableline\tableline
\multicolumn{1}{l}{Input Data} &
\multicolumn{1}{c}{}\\
\tableline
$\delta v$ = 500 km s$^{-1}$  &  \\
~~~satellite sample          &  $-1.21 \pm 0.09$   \\
~~~clustered random points   & $-1.46 \pm 0.11$  \\
~~~nearby points ($\Delta R_{\rm corr}$ = 1 - 2)    & $-1.58 \pm 0.11$ \\
$\delta v$ = 1000 km s$^{-1}$     & \\
~~~satellite sample          &  $-1.18 \pm 0.09$ \\
~~~clustered random points   & $-1.55 \pm 0.11$   \\
~~~nearby points ($\Delta R_{\rm corr}$ = 1 - 2)    & $-1.65 \pm 0.12$  \\
\tableline
\end{tabular}
\end{center}
\end{table}

The statistics of primaries and satellites for both samples are shown
in Figure \ref{fig:hist}.  The number of possible satellites found in
the volume-limited samples is small.  For the range $-23 < M_{r} <
-20$, there are 871 primary galaxies and 336 objects in the satellite
sample with projected radii greater than the minimum separation and
less than 0.5 $h^{-1}$ Mpc.  
%Change here;  comment 5.
For the $\delta v$ = 1000 km s$^{-1}$ sample, which uses a slightly 
larger minimum separation and limiting magnitude, 
there are 357 galaxies.  The volume-limited samples are summarized in
Table \ref{tab:select_sdss}.

The satellite sample and the nearby points interloper subtracted
results are shown in Figure \ref{fig:fit_data} for the velocity
criterion of $\delta v$ = 500 km s$^{-1}$, in bins of 70 $h^{-1}$ kpc,
starting from the minimum separation of $32.9 h^{-1}$ kpc.  The
results are similar to those in the simulations; the satellite sample
distribution is shallower than the interloper subtracted samples.  In
addition, the nearby points method distribution is steeper than that
of the clustered random points method.

We fit the radial profile with a power law, with the results for the
slope marginalized over the amplitude of the power law shown in Table
\ref{tab:slope_all}.  The slope of the best-fit power law for the
satellite sample is $-1.21 \pm 0.09$ in the sample with $\delta v$ =
500 km s$^{-1}$, with very similar results for the $\delta v$ = 1000 km s$^{-1}$
sample.  The clustered random points method finds a slope of $-1.46
\pm 0.11$ and the nearby points methods finds a slope of $-1.58 \pm
0.11$ in the $\delta v$ = 500 km s$^{-1}$ sample.  The $\delta v$ = 1000
km s$^{-1}$ sample shows steeper fits, with slopes of $-1.55 \pm 0.11$ and
$\-1.65 \pm 0.12$, respectively.  The systematic differences between
the best-fit slopes of the clustered random points and nearby points
is $\sim 0.1$, consistent with the result found in the simulations.
In the simulations, the bias found in the interloper subtraction
methods was $\sim 0.1$ for the nearby points method and $\sim 0.2$ for
the clustered random points method, which would imply that the slope
of the true satellite distribution is $\alpha=\alpha_{\rm
  est}-\Delta\alpha_{\rm bias} \approx -1.7$.

The marginalized errors suggest that slopes of the satellite samples
and the interloper subtracted samples are significantly different.  We
illustrate this point in Fig. \ref{fig:fit_data}, where we plot all
the points at projected radii smaller than 0.5$h^{-1}$ Mpc for the
$\delta v$ = 500 km s$^{-1}$, satellite sample and the results of the nearby points method.
Here the best-fit power-law is plotted in the top panel, where the
best-fit slope of the satellite sample is $\alpha$ = -1.22 and the
best-fit slope of the interloper subtracted sample is $\alpha =
$-1.58.  The bottom panel shows the confidence regions for the two
fits, where the slopes of the two distributions do not overlap within
the 99\% confidence intervals.

%---------------------------------
\subsection{Completeness test}
\label{sec:completeness}
%---------------------------------

\begin{figure}[h]
\epsscale{1.1}
\plotone{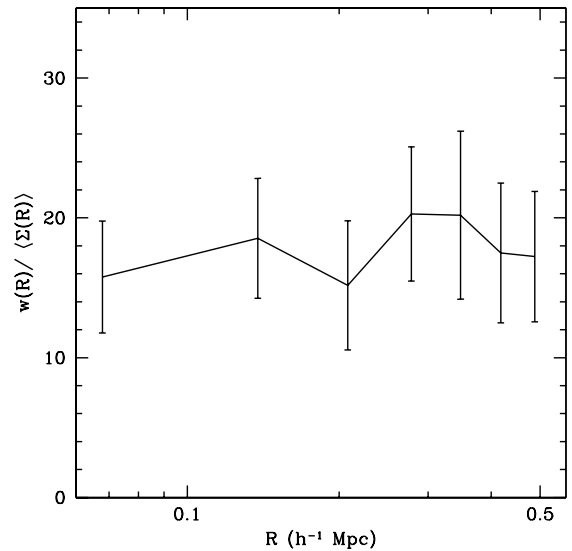}
\caption{The ratio of the cross-correlation function, $w(R)$, 
to the projected, interloper subtracted surface density of 
true satellites estimated using the random points method 
for the volume-limited sample with $\delta v$ = 500 km s$^{-1}$. \label{fig:wp}}
\end{figure}

Although we have selected our sample carefully to avoid spurious
results, we still need to assess to what extent our results
might be biased by sample incompleteness, caused by, for example, holes
in the survey or edge effects.  We can test this by calculating for
our primary$+$ satellite sample a statistic analogous to the
projected cross-correlation function:
\begin{equation}
w(R) = \frac{\langle N_{\rm sat}(R)\rangle}
{\langle N_{\rm int}(R)\rangle} - 1,
\end{equation}
where $\langle N_{\rm sat}(R) \rangle=\langle N_{\rm truesat}(R)
\rangle +\langle N_{\rm int}(R)\rangle$ and $\langle N_{\rm
  int}(R)\rangle$ is the estimated number of interlopers at separation
$R$ in a sample, and $\langle N_{\rm truesat}(R) \rangle$ is the
corresponding estimated average projected number of the true
satellites.  For the purposes of this test $\langle N_{\rm
  int}\rangle$ is estimated using the random points method (i.e.,
assuming uniform projected density of interlopers).

The function $w(R)$ can be compared to the surface density of the
true satellite galaxies, $\Sigma_{\rm truesat}(R)$, estimated using the
random points method.  If the estimate of $\Sigma_{\rm truesat}(R)$ is affected by
incompleteness, the functions $w(R)$ and $\Sigma_{\rm truesat}(R)$ should
have different shapes because $w(R)$ is defined as a ratio of
quantities equally affected by area incompleteness, which should
cancel out the effect.  $\Sigma_{\rm truesat}(R)$, on the other hand, will
be affected.  Conversely, $w(R)$ and $\Sigma_{\rm truesat}(R)$ should have
the same shape if effects of incompleteness on $\Sigma_{\rm truesat}(R)$
are negligible. This is because for the random points interloper
subtraction: $\Sigma_{\rm truesat}=(\langle N_{\rm sat}\rangle/\langle N_{\rm int}\rangle -1)\Sigma_{\rm int}=w\Sigma_{\rm int}$, where $\Sigma_{\rm int}(R)=\rm const$.

Figure \ref{fig:wp} shows the ratio of the surface density profile of
satellites estimated for our volume-limited sample using the random
points method and function $w(R)$, computed for the same sample.  Over
the projected radii test, the ratio is consistent with a constant.  In
addition, the marginalized best-fit slopes of $\Sigma_{\rm
  truesat}$ and $w(R)$ agree within statistical errors: $\alpha =
-1.33 \pm 0.09$ and $\alpha = -1.41 \pm 0.06$, respectively.  We
therefore conclude that our measurements of the surface density
profiles of satellites are not significantly affected by area
incompleteness.

%---------------------------------
\subsection{Flux-limited samples}
\label{sec:flux}
%---------------------------------

\begin{figure}[h]
\epsscale{1.}
\plotone{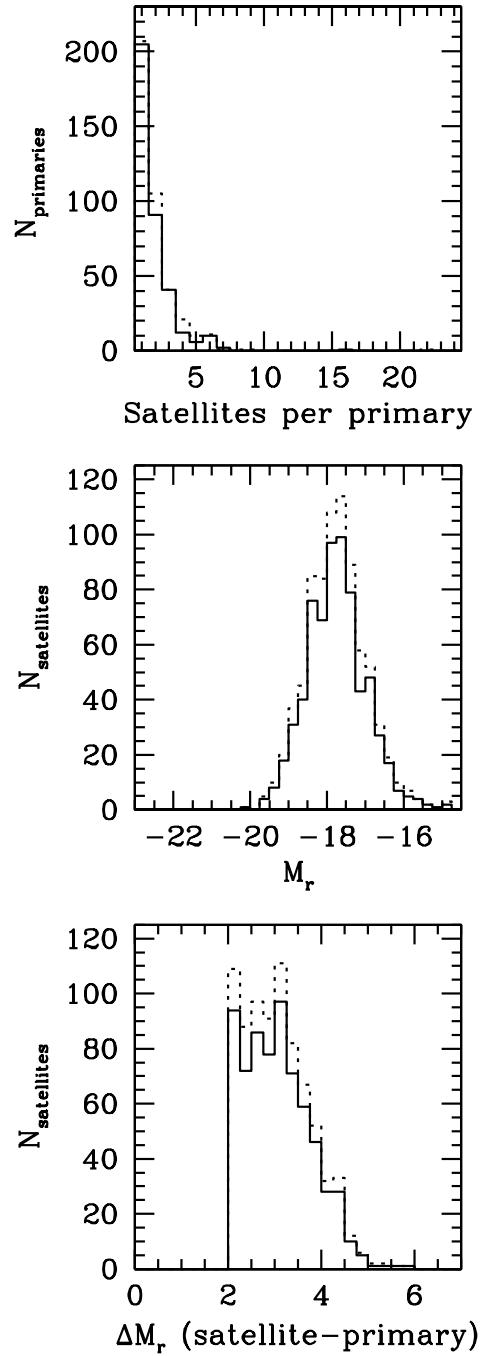}
%/home/jchen/satgals/data/erin_sdsscat/DAT/hist_big.ps
\caption{Statistics of the satellite samples for a flux-limited sample.  The solid line shows the results for the $\delta v$ = 500 km s$^{-1}$ criterion, while the dotted line shows the $\delta v$ = 1000 km s$^{-1}$ criterion.  {\it Top:}  The number of satellites per primary for primaries with at least one satellite.  {\it Center:}  The $r$-band magnitude histogram for satellites.  {\it Bottom:}  The magnitude differences between satellites and primaries.  \label{fig:hist_big}}
\end{figure}

%\begin{figure}[h]
%\epsscale{1.}
%\plotone{all_big.ps}
%%/home/jchen/satgals/data/erin_sdsscat/DAT/all_big.ps
%\caption{The secondary sample compared to interloper subtracted samples for the flux-limited sample using the fiducial criteria.  The projected radial number density profile for the secondary sample is the solid line, the interloper subtracted profile for clustered random points is the dashed line, and nearby points is the dotted line.  \label{fig:all_big}}
%\end{figure}

\begin{figure}[h]
\epsscale{1.}
\plotone{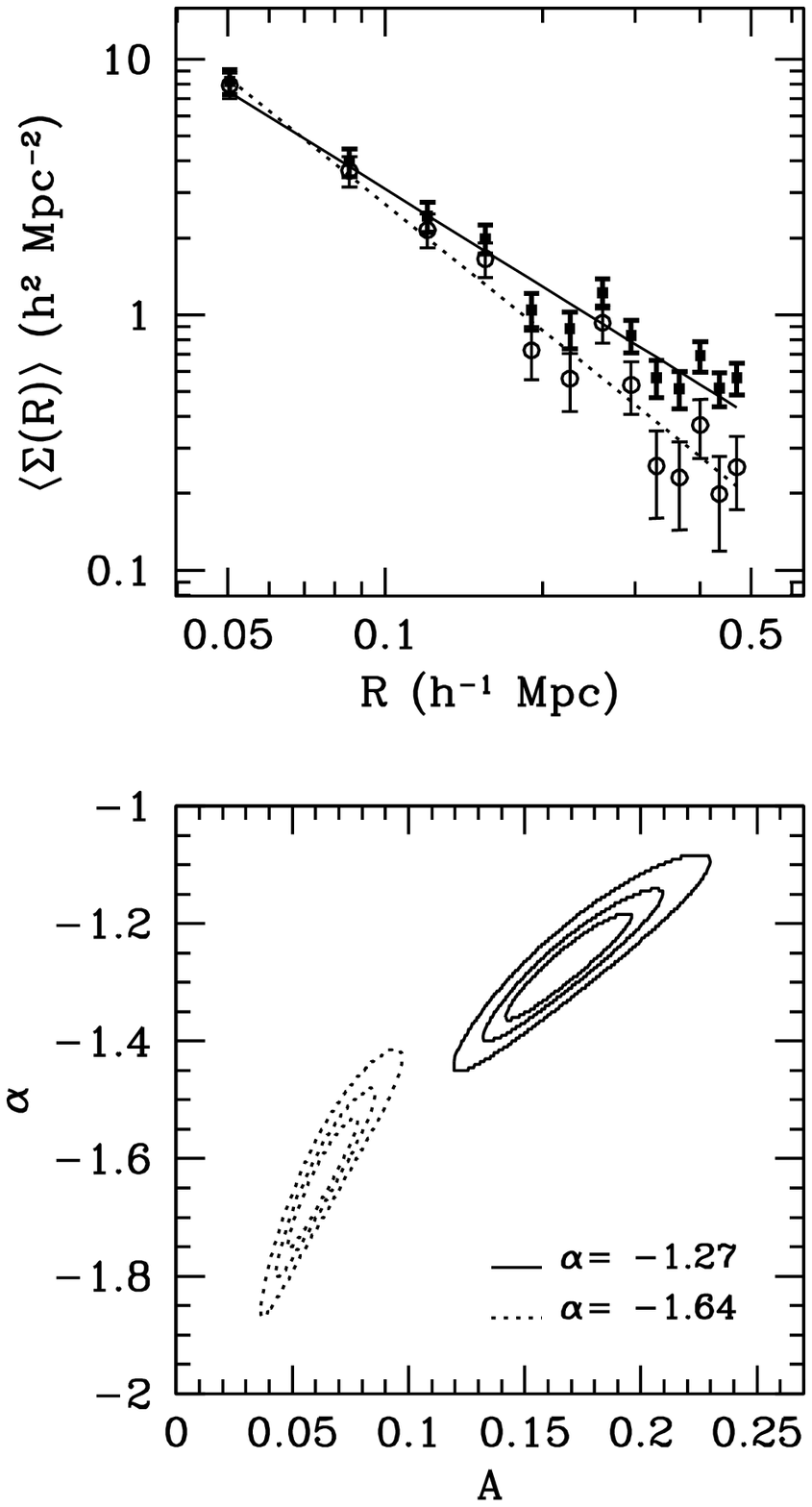}
%/home/jchen/satgals/data/erin_sdsscat/DAT/fit_data_big.ps
%Changed figure;  there was an error in it previously.
\caption{The satellite sample compared to interloper subtracted samples for the flux-limited sample with velocity criterion $\delta v$ = 500 km s$^{-1}$.  {\it Top:}  The best-fit power-law for the satellite sample (solid) and the interloper subtracted profile for the $\Delta R_{\rm corr}$ = 1 - 2 $h^{-1}$ Mpc nearby points (dotted).  The data is shown with error bars:  satellite sample (square points, thick Poisson error bars) and nearby points (open circles, thin error bars).  {\it Bottom:} The 68\%, 90\%, and 99\% confidence intervals for the satellite sample (solid lines) and the nearby points sample (dotted lines).  
%Change here;  comment 7.
The contours are labeled by the unmarginalized best-fit slope of each profile. \label{fig:fit_data_big}}
\end{figure}

\begin{table}[h]
\begin{center}
\caption{Estimated Power-Law Slopes for the Flux-Limited Sample\label{tab:slope_all_big}}
\begin{tabular}{lc}
\tableline\tableline
\multicolumn{1}{l}{Input Data} &
\multicolumn{1}{c}{}\\
\tableline
$\delta v$ = 500 km s$^{-1}$ \\
~~~satellite sample       &     $-1.27 \pm 0.06$  \\
~~~clustered random points      & $-1.52 \pm 0.07$ \\
~~~nearby points ($\Delta R_{\rm corr}$ = 1 - 2) ~~~    & $-1.64 \pm 0.07$  \\
$\delta v$ = 1000 km s$^{-1}$ \\
~~~satellite sample         &  $-1.17 \pm 0.06$ \\
~~~clustered random points       & $-1.56 \pm 0.07$ \\
~~~nearby points ($\Delta R_{\rm corr}$ = 1 - 2) ~~~    & $-1.66 \pm 0.08$ \\
\tableline
\end{tabular}
\end{center}
\end{table}

We could significantly increase the
number of objects in our samples by eliminating the limiting absolute magnitude and increasing the maximum depth of the sample.  Increasing the depth, however, will increase the minimum separation of the sample as dictated by the fiber collision problem.  We, then, only eliminate the magnitude limit to create a flux-limited sample.   For these flux-limited samples, we
apply the same isolation and satellite criteria as the volume-limited
sample.  
%Change here; comment 5.
For the satellite sample with $\delta v$ = 500 km s$^{-1}$, we have 678 galaxies.  For the larger $\delta v$ = 1000 km s$^{-1}$ sample, with a larger minimum separation and limiting magnitude, there are 
786 galaxies.  The flux-limited samples are summarized in Table \ref{tab:select_sdss}.  Statistics of the satellite samples in the flux-limited set are plotted in Fig.
\ref{fig:hist_big}, while the satellite sample and the nearby points interloper
subtracted results with $\delta v$ = 500 km s$^{-1}$ are plotted in Figure \ref{fig:fit_data_big} for
bins of 35$h^{-1}$ kpc starting at the minimum separation of 32.9 $h^{-1}$ kpc.  In
addition, the marginalized best-fit slopes are listed in Table
\ref{tab:slope_all_big}.

The flux-limited results are consistent with those of the
volume-limited sample.  The clustered random points and nearby points methods for
interloper removal once again show steeper profiles than the satellite
sample.  We also find best-fit power-laws that are similar to the
volume-limited sample for the satellite sample ($\alpha = -1.27$) and
nearby points method ($\alpha = -1.64$) in the top panel of Fig.
\ref{fig:fit_data_big}.  The bottom panel shows the confidence regions
for the two fits, where the slopes of the two distributions do not
overlap within the 99\% confidence intervals.

The marginalized best-fit power-law slope values in Table
\ref{tab:slope_all_big} show similar results to those found in the
volume-limited sample.  The satellite sample with $\delta v$ = 500 km s$^{-1}$ has a
somewhat steeper slope than the $\delta v$ = 1000 km s$^{-1}$ sample, $-1.27 \pm 0.06$ to $-1.17
\pm 0.06$.  In addition, the clustered random points and nearby points
methods produce very consistent results in both samples, $\alpha =
-1.52 \pm 0.07$ and $\alpha = -1.64 \pm 0.07$ for the $\delta v$ = 500 km s$^{-1}$ sample.

The similarity of the results for the volume-limited and flux-limited
samples suggest that the flux-limited data do not induce any bias to
the sample while increasing the statistical significance of the
results.

%--------------------------------
\subsection{Trends with luminosity}
%--------------------------------

\begin{figure}[h]
\epsscale{1.}
\plotone{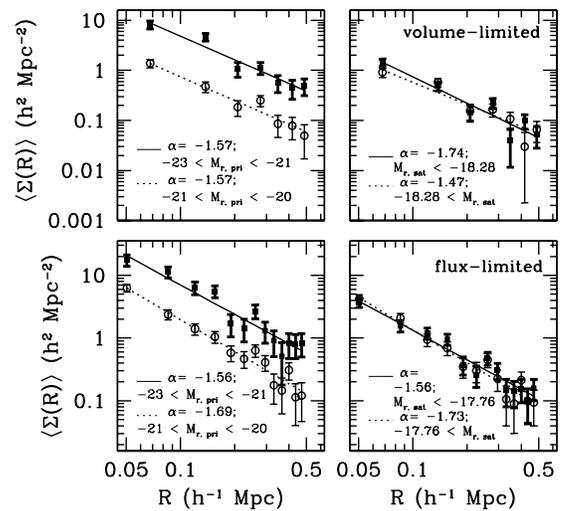}
%/home/jchen/satgals/data/erin_sdsscat/DAT/fit_data_pricut_satcut.ps
\caption{The left-hand panels show the results of the nearby points method for samples split by bright and faint primaries.  The right-hand panels shows the results when samples are split into bright and faint satellites.  In all cases, the 'bright' data is shown in square points, with thick error bars and the 'faint' data is shown in open circles with thin error bars.  In addition, the 
%Change here;  comment 7.
unmarginalized 
best-fit power-law for the 'bright' data is shown with a solid line and a dotted line for the 'faint' data. The top panels show results from the volume-limited sample and the bottom panels show the results from the flux-limited sample.  \label{fig:fit_data_pricut_satcut}}
\end{figure}

We now test the projected radial distribution for differences between
bright and faint primary galaxies and bright and faint satellites (see
Figure \ref{fig:fit_data_pricut_satcut}) for the nearby points method and $\delta v$ = 500 km s$^{-1}$.
We first split the primary sample into two, a bright sample of $-23 <
M_{r} < -21$ with 125 primaries and a faint sample of $-21 < M_{r} <
-20$ with 746 primaries.  The volume-limited set has 161 possible
satellites in the bright primaries sample and 171 in the faint
primaries sample.  The flux-limited set has 225 possible satellites in
the bright primaries sample and 297 in the faint primaries sample.
The left-hand panels of Figure \ref{fig:fit_data_pricut_satcut} show
that, in all cases, the amplitude of the projected number density
profile is always larger for the bright primary samples.  The bright
and faint primary samples show consistent slopes. The slopes
marginalized over the normalization for the bright 
volume- and flux-limited samples are $-1.56 \pm 0.14$
and $-1.55 \pm 0.11$, respectively, while the 
corresponding slopes for the faint samples are
 $-1.57 \pm 0.17$ and $-1.68 \pm 0.10$. We thus do not find 
a significant dependence of the concentration of the 
radial profiles on the luminosity of the primary galaxies
in the luminosity range probed. 

Next, we split the satellite sample into samples of bright and faint
satellites at the median absolute magnitude of each satellite sample.
The volume-limited set is split at $M_{r}$ = -18.28, while the
flux-limited at $M_{r}$ = -17.76.  The luminosity range of satellite
galaxies probed is not large, with the medians of the satellite
samples $\sim 0.1 L_{*}$.  The right-hand panels of Fig.
\ref{fig:fit_data_pricut_satcut} show that the best-fit power-law
slopes, marginalized over the normalization, for the bright and faint
satellites are consistent: $-1.72 \pm 0.16$ and $-1.45 \pm 0.16$,
respectively, for the the volume-limited set, and $-1.55 \pm 0.11$ and
$1.72 \pm 0.11$ for the flux-limited set. 

%-----------------------------------
\subsection{Putting everything together: comparisons with 
$\Lambda$CDM expectations}
\label{sec:everything}
%-----------------------------------

In this section we compare the observed projected density profile of
satellites in our SDSS sample to mock satellite
samples derived from high-resolution simulations of the concordance
$\Lambda$CDM model.  For this comparison, we compare to 
catalogs that are representative of subhalo, satellite galaxy, and dark matter
distributions, interesting reference points to compare to the
observational results.  In the first catalog, we use
samples of isolated halos and subhalos selected simply using their
maximum circular velocity, $V_{\rm max}$.  The only criterion we use
for subhalos is $V_{\rm max}> 100\ {\rm km\,s^{-1}}$ to avoid possible
biases due to effects of resolution for smaller subhalos.  We use two additional samples to probe the subhalo distribution given two different ways of selecting subhalos.  In the first,
galaxies and satellite galaxies are associated with halos and subhalos, and
selection is based on the $r$-band galaxy luminosities assigned in
such a way that the observed luminosity function of galaxies, their
clustering, and galaxy-mass correlations are well reproduced
\citep[see \S~\ref{sec:testing} and ][for details]{tasitsiomi_etal04}.
The selection criteria of the primary galaxies and satellites mimic
those applied to the SDSS samples.  This sample is analogous to Test Sample 3 (see \S~\ref{sec:testing}), except that our sample is taken at $z=0$. 
\citet*{conroy_etal05b} present a very promising modification to the
luminosity assignment scheme discussed which, coupled with 
high-resolution dissipationless simulations, reproduces the small-scale
galaxy clustering and its luminosity dependence observed in SDSS 
remarkably well.  This scheme assigns $r$-band galaxy luminosities for subhalos not on the $V_{\rm max}$ at $z=0$ but using the $V_{\rm max}$ at the epoch of accretion for the subhalo.  Given that the baryonic components of subhalos should be more resistant to the physical processes that evolve the properties of subhalos as they fall into their parent halo, this sample should better reflect the distribution of satellite galaxies.  Finally, we use the same primaries as for the subhalo sample, but the
satellite sample is now constructed by randomly sampling DM particles
surrounding primaries.

\begin{figure}[t]
\epsscale{1.2}
\plotone{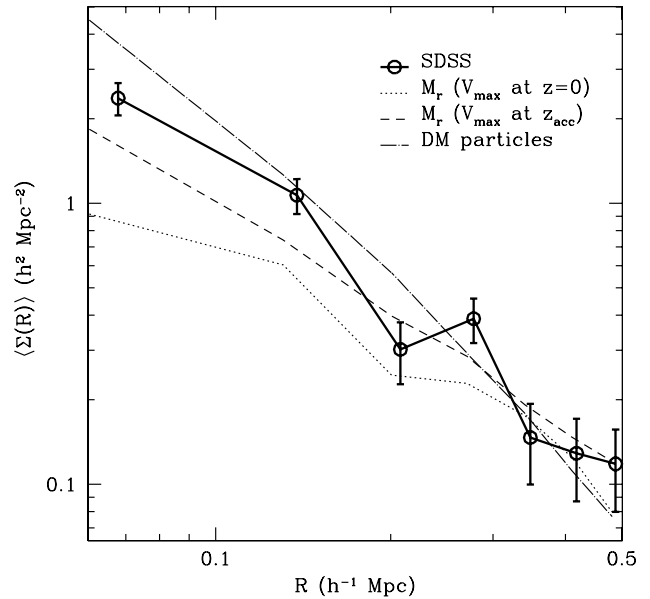}
\caption{Comparison of the projected number density of satellites
  in the volume-limited SDSS sample ({\it open circles with error bars
    connected by the solid line}) and in the three mock satellite
  simulation-derived samples (see text for details). The {\it
    dot-dashed} line represents dark matter distribution around
  primary halos in simulations, while {\it dotted} and {\it dashed}
  lines show satellite profiles using two different ways of assigning luminosities to subhalos (see text). }
\label{fig:sigmasum}
\end{figure}

Figure~\ref{fig:sigmasum} shows the interloper-corrected profile for
our volume-limited SDSS sample along with the profiles for the
true satellites distribution for satellites assigned luminosities by $V_{\rm max}$ at $z=0$ and $V_{\rm max}$ at accretion and the sample of dark
matter particles around halos with $V_{\rm max}$ values in the range
$200-300\ {\rm km\,s^{-1}}$.  The normalization of the dark matter particle 
profile is set arbitrarily in the plot.  The figure shows that the shape of the
profiles for all three simulation-derived samples is similar to that of the
SDSS sample.  The distribution for satellites assigned luminosities by $V_{\rm max}$ at $z=0$, however, is shallower than the observed distribution.  The distribution for satellites assigned luminosities by $V_{\rm max}$ at accretion also appears somewhat shallower, but less so than for the other luminosity assignment.  The dark matter particles, on the other hand, have a somewhat steeper radial profile than the observed satellites.

\begin{figure}[t]
\epsscale{1.}
\plotone{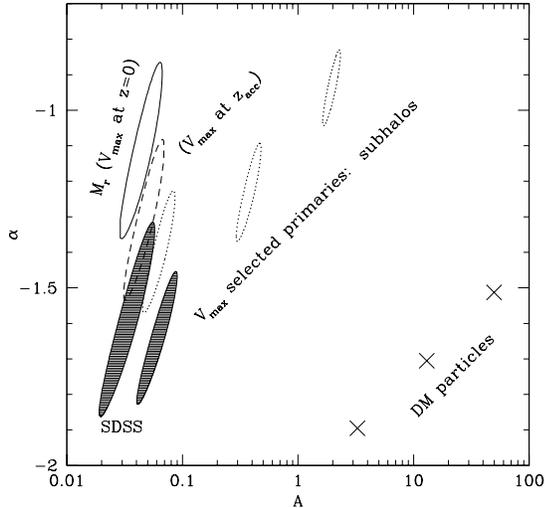}
%/home/jchen/satgals/Results/new/compar_samples/sum_plot/sum_test.ps
\caption{The 95.4\% confidence intervals for the best-fit power-laws of the interloper subtracted data and for the true satellite distributions of several samples drawn from the simulations.  The two filled contours represent the volume-limited sample (on the left) and the flux-limited sample (on the right).  The {\it solid} and {\it dashed} contours show satellite samples using two different luminosity assignment schemes, as discussed in the text.  Samples are drawn from $V_{\rm max}$ selected primary samples for subhalos with $V_{\rm max} >$ 100 km s$^{-1}$ (dotted contour lines) and for DM particles (crosses), with the $V_{\rm max}$ ranges of 200-300, 300-400, and 400-600 km s$^{-1}$ showing increasingly shallower slopes and larger amplitudes.  The normalization, $A$, for 
samples of DM particles is not meaningful, as the samples are drawn from arbitrary fractions of DM particles around host halos.}
\label{fig:alphadatatheory}
\end{figure}

Figure~\ref{fig:alphadatatheory} compares the best fit slopes for the
full range of radii for the different simulation-derived samples discussed above with the corresponding best fit slope for the
radial profile of the SDSS satellites.  We present
radial distributions of subhalos and dark matter particles for the host
halos in three ranges of $V_{\rm max}$ to illustrate the dependence of
the satellite distribution of the parent system mass.  For the $V_{\rm max}$-selected subhalo catalog for primary
systems with $V_{\rm max}=200-300\ {\rm km\,s^{-1}}$, the slope and
normalization of the distribution are in reasonable agreement with
the data at the 2$\sigma$ level. The slope is somewhat shallower
($\alpha \approx -1.3-1.4$) than observed, but overlap with the latter
within the uncertainties. The slope of the 
corresponding catalog in which satellites are selected from DM particles, 
exhibit
a somewhat steeper slope ($\alpha\approx -1.8$) than observed for the
SDSS galaxies, although the difference is not statistically
significant.  Note that the normalization, $A$, for this sample is not
meaningful, while it is meaningful for the subhalo sample.  
%Change here in ref. to comment 9;  added text.
For primary
systems with larger $V_{\rm max}$ ranges, the subhalo radial
disribution becomes
progressively shallower than the data because the fixed projected 
radial range used probes different radii with respect to the halo 
virial radii.  
Most of the primaries in our sample should correspond to halos with
$V_{\rm max}<300\ \rm km\, s^{-1}$, so it is not surprising that the data 
matches the subhalo distribution for primaries with 
$V_{\rm max}=200-300\ {\rm km\,s^{-1}}$ the best.  For primaries 
of this size, the virial radius is expected to significantly 
smaller than 500 $h^{-1}$ kpc, the maximum projected radius 
measured.  Simulations, however, show bound subhalos at several 
times the virial radii of the host halos.  

Figure~\ref{fig:alphadatatheory} also compares the best fit slopes for
the SDSS to satellite samples with different luminosity assignments.
Both assignments are in agreement with the best fit slope of the SDSS
volume-limited sample at the 2$\sigma$ level.  Assigning luminosities
based on a subhalo mass at accretion, however, results in a steeper
slope than assigning luminosities based on a subhalo mass at $z=0$
and is more consistent with the data.

As previously discussed, the slopes of satellite surface density 
profiles around SDSS galaxies
are in between those measured for subhalos and those for dark matter
distributions in simulations of the concordance cosmology, and 
the measured slopes are closer to those measured for dark matter
distribution.  In addition, the bias in the interloper subtraction
methods suggest that the true satellite distribution is somewhat
steeper than our best fit value and even more compatible with the dark
matter distribution.  Our results are in qualitative agreement with
the recent observational studies on the radial distribution of
galaxies in groups and clusters
\citep[e.g.,][]{lin_etal04,hansen_etal05,collister_lahav05,yang_etal05,coil_etal05},
which find concentrations of galaxy radial profiles somewhat lower
than the concentrations expected for the matter distribution of
their parent halos. However, here we seem to find that the difference between 
the DM distribution and the satellite distribution 
is not as large in galaxy-sized systems compared to groups and clusters.

\begin{table}[t]
\begin{center}
\caption{Estimated Power-Law Slopes for Different Samples\\ of True Satellites\label{tab:simslopes}
}
\begin{tabular}{lrr}
\tableline\tableline
\multicolumn{1}{l}{Primary$+$satellite sample} &
\multicolumn{1}{r}{$R_{\rm min}<R< 0.5$} &
\multicolumn{1}{r}{~~$0.1 < R< 0.5$}\\
\tableline
%Test Sample 2 & $-1.815 \pm 0.007$ & $-2.051 \pm 0.016$ \\
%Change here; comment 8.
SDSS (volume) & $-1.58 \pm 0.11$ & $-1.77 \pm 0.23$ \\
SDSS (flux) & $-1.64 \pm 0.07$ & $-1.72 \pm 0.17$\\
\tableline
%Test Sample 3 & $-1.34 \pm 0.12$ & $-1.53 \pm 0.23$ \\
$M_{r}$ ($V_{\rm max}$ at $z=0$)  & $-1.12 \pm 0.10$ & $-1.36 \pm 0.18$ \\
$M_{r}$ ($V_{\rm max}$ at $z_{\rm acc}$)  & $-1.31 \pm 0.09$ & $-1.38 \pm 0.17$ \\
$V_{\rm max}$ = 200-300 (DM) & $-1.895 \pm 0.010 $ & $-2.111 \pm 0.020 $ \\
$V_{\rm max}$ = 200-300 (subs) & $-1.40 \pm 0.07$ & $-1.65 \pm 0.13$\\
$V_{\rm max}$ = 300-400 (DM) & $-1.705 \pm 0.009$ & $-1.864 \pm 0.018$ \\
$V_{\rm max}$ = 300-400 (subs) & $-1.23 \pm 0.06$ & $-1.65 \pm 0.11$ \\
$V_{\rm max}$ = 400-600 (DM) & $-1.513 \pm 0.008$ & $-1.674 \pm 0.015$ \\
$V_{\rm max}$ = 400-600 (subs) & $-0.94 \pm 0.04$ & $-1.26 \pm 0.07$ \\
\tableline
\end{tabular}
\end{center}
{\small Note -- Samples using isolated halos chosen by $V_{\rm max}$ and subhalos assigned a $r$-band luminosity use binning of 70$h^{-1}$~kpc, with first bin starting at $R_{\rm min} = 25.6h^{-1}$ kpc. 
Subhalos in these samples are selected to have
circular velocities $V_{\rm max}\geq 100\ \rm km\ s^{-1}$.  Data samples use binning previously described in the text.}
\end{table}
 
Several previous studies which
considered satellite distribution in {\it cluster-sized} halos showed that subhalos appear to have more extended and
shallower radial distributions compared to that of dark matter 
\citep{ghigna_etal98,colin_etal99,ghigna_etal00,springel_etal01,delucia_etal04,diemand_etal04,gao_etal04,nagai_kravtsov05}.
Tidal evolution and merging modify the subhalo profile, especially
within inner $\sim 50\%$ of the virial radius, primarily because they
modify properties of subhalos, such as its bound mass or circular
velocity \citep{nagai_kravtsov05}. For galaxy-sized systems in our simulations, then, we
would thus expect subhalo profile to be significantly flattened at
$R\lesssim 100 h^{-1}$~kpc.  
Table~\ref{tab:simslopes} presents the best-fit slopes for the
full range of radii and for the fits restricted to radii between $0.1$
and $0.5h^{-1}$~Mpc for all the data and simulation samples discussed.  The table shows that the best-fit
power-laws for bins of $R> 100 h^{-1}$ kpc are steeper than for those including smaller projected radii.  This is generically true, but in particular subhalo distributions seem to steepen more than the DM distributions, suggesting that subhalo distributions are flattened within $\sim$50\% of the virial radius as in cluster-sized halos.  For example, for subhalos assigned luminosities at $z=0$, the slope for the restricted set of radii is $-1.36\pm 0.18$ -- $\approx 0.2$ steeper than the fit 
using all data points
-- and the slope steepens further if we constrain the fit to even
larger radii, where the best fit slope values of DM and subhalo
distributions agree within error bars (note, however, that at these
radii the fit errors become larger).  On the other hand, the slope for subhalos assigned luminosities at the epoch of accretion and the observational results steepen less, as would be expected given that the baryonic components of subhalos should be more resistant tidal evolution and merging.  We attribute some of the differences in slopes and 
shallower slopes for our simulation satellite samples to tidal
evolution effects, and to the differences in object selection in
simulation and SDSS samples \citep{nagai_kravtsov05}. A better
understanding of the differences between selection criteria is
necessary for more stringent tests of the theoretical predictions.

%-----------------------------------
\section{Discussion and Conclusions}
\label{sec:conclusions}
%-----------------------------------

Modern large galaxy redshift surveys allow one to study the distribution
of satellites around galaxies and clusters with unprecedented
statistical power, while controlling biases and completeness in a
systematic way.  In addition, redshift information can be used to
select galaxies from relatively non-crowded environments and to
account for interloper contamination in a rigorous way.  Cosmological
simulations are also sufficiently mature and allow systematic tests of
the interloper subtraction algorithms.  Galaxy and satellite samples,
for instance, can be constructed to mimic observational selection
criteria. Examples of studies using such surveys are constraints on
the DM halos of galaxies from satellite kinematics
\citep{mckay_etal02,prada_etal03,brainerd05b} and the anisotropy of
the distribution of satellite galaxies
\citep{sales_lambas04,brainerd05}.

In this work, we use the SDSS spectroscopic survey to estimate the
projected radial distribution of satellites around isolated primaries.
We use areas of the survey which are at least 90\% complete and check
for the effects of incompleteness by comparing the surface density
profile and conclude that our results are not affected by the (small)
incompleteness of the sample.  We construct samples of primary and
satellite galaxies with isolation criteria similar to those used by
\citet{prada_etal03}. We use high-resolution cosmological simulations
of the concordance $\Lambda$CDM cosmology to develop and carefully test
new methods of correcting for interloper contamination. 
Our main results and conclusions can be summarized as follows. 

\begin{itemize}
  
\item[1.] Using mock galaxy catalogs derived from high-resolution
  cosmological simulations, we show that interlopers can significantly
  bias the shape of the projected surface density profile of faint
  satellites around bright galaxies, making it shallower (biasing the
  power-law slope $\alpha$ of the radial profile, $\Sigma(R)\propto R^{\alpha}$, by 
  $\Delta\alpha_{\rm bias}\gtrsim 0.5$). We also
  show that the most straightforward methods do not correct interloper
  contamination properly.  For example, the random points
  method, which assumes uniform distribution of interlopers in space,
  underestimates the fraction of interlopers in the satellite sample
  by oversampling voids compared to clustered areas where most
  galaxies in the sample reside.
  
\item[2.] We develop two new methods to account for the interloper
  contamination: the clustered random points method and the nearby
  points method, variants of the random points method, designed to
  sample environments similar to those of the clustered galaxies in
  the observed samples.  Tests on the mock samples show that the
  methods perform consistently well, reducing the interloper bias in
  the best-fit power-law slopes of the satellite profiles 
  to only $\Delta\alpha_{\rm bias}\approx 0.1$ for the nearby points method.
  
\item[3.]  We apply these methods in our analyses of the volume- and
  flux-limited SDSS spectroscopic samples. The best fit power-law
  slope for the volume-limited SDSS satellite sample, after interloper
  contamination correction, is $-1.58 \pm 0.11$ in the range of
  projected separations of $32.9<R<500h^{-1}$~kpc.  Note that we
  estimated a systematic bias in the derived slope of
  $\Delta\alpha_{\rm bias}\approx 0.1$ for nearby points interloper
  subtraction method, which implies that the true slope of the SDSS
  satellites may be $\alpha=\alpha_{\rm raw}-\Delta\alpha_{\rm bias}
  \approx -1.7$.  We find similar values of the best fit slopes for
  the flux-limited samples, and for samples of primary galaxies in
  different absolute magnitude ranges. We thus do not find evidence
  for the dependence of the shape of satellite radial distribution on
  the luminosity of their host galaxy.

\item[4.] Comparison of the observed radial distribution of the SDSS 
satellites to the distribution of subhalos and dark matter around 
galactic halos in dissipationless $\Lambda$CDM simulations shows that 
the slope of the SDSS satellite radial profile
is in between those measured for subhalos and for dark matter (closer
to dark matter).  Subhalos thus appear to have more extended and
shallower radial distributions than the observed 
satellites. The dark matter distribution is somewhat steeper 
than the observed satellite profiles, but the difference is not
statistically significant. 
\end{itemize}

Recently, \citet{vandenbosch_etal05} and \citet{sales_lambas05}
studied the projected radial distribution of satellite galaxies around
isolated galaxies using the Two Degree Field Galaxy Redshift Survey
(2dFGRS).  
% Changed sentence below in reference to comment 10.
\citet{vandenbosch_etal05} used mock galaxy redshift samples derived
from large cosmological simulations to develop an iterative method of 
interloper rejection for the 2dFGRS and found that the data is generally consistent with the 
dark matter profile at large projected radii, but concluded that 
incompleteness of close pairs in the survey prevent strong constraints.
\citet{sales_lambas05} used isolation
criteria for the primary galaxies, enforcing that any neighbor within
a region of 700 $h^{-1}$ kpc and $\Delta V$ = 1000 km s$^{-1}$ must be at
least 1 magnitude fainter.  Satellites were assumed to be any object
within projected separation of $500$~kpc and $\Delta V = 500\ {\rm
  km\, s^{-1}}$ that is at least 2 magnitudes fainter than the host
galaxy. \citet{sales_lambas05} estimated the effect of close-pair
incompleteness by deriving a control sample of projected satellites
with velocity difference of $2000 < \Delta V < 10,000\ \,{\rm km\,
  s^{-1}}$. At $R \gtrsim$ 20 $h^{-1}$ kpc, where the number density
profile of the control sample is flat, the satellite samples were
considered complete.  For the total sample, \citet{sales_lambas05}
quoted the best-fit power-law of slope $\alpha = -0.96 \pm 0.03$. This
value, however, was derived without applying correction for interloper
contamination.  Note that we derive significantly steeper profiles of
satellites compared to those of \citet{sales_lambas05}, which we
attribute to the rigorous correction for interlopers and to possible
fiber-collision bias in the 2dFGRS.

The results presented in this study provide interesting hints of the
possible differences between observed satellite distributions and the
expected distribution of subhalos in their parent halos. However, the
statistical errors are still rather large. Significant improvements in
the statistics are needed to address this question further. Larger statistics
would also allow us to go beyond the average profiles and study the
distribution of satellites as a function of satellite (e.g.,
luminosity and color) and host galaxy properties.
    
\acknowledgements

We would like to thank Risa Wechsler for her help with the luminosity
assignment for the halo catalogs and Charlie Conroy for a careful
reading of the manuscript and useful comments. 
This research was carried out at the University of Chicago, Kavli
Institute for Cosmological Physics and was supported (in part) by
grant NSF PHY-0114422. KICP is an NSF Physics Frontier Center. JC and
AVK were supported by the National Science Foundation (NSF) under
grants No.  AST-0206216, AST-0239759 and AST-0507666, and by NASA
through grant NAG5-13274. Cosmological simulations used in this
analysis were performed on the IBM RS/6000 SP3 system ({\tt seaborg})
at the National Energy Research Scientific Computing Center (NERSC).
This work has made use of the NASA Astrophysics Data System. 

Funding for the creation and distribution of the SDSS Archive has been provided by the Alfred P. Sloan Foundation, the Participating Institutions, the National Aeronautics and Space Administration, the National Science Foundation, the U.S. Department of Energy, the Japanese Monbukagakusho, and the Max Planck Society. The SDSS Web site is http://www.sdss.org/. 

The SDSS is managed by the Astrophysical Research Consortium (ARC) for the Participating Institutions. The Participating Institutions are The University of Chicago, Fermilab, the Institute for Advanced Study, the Japan Participation Group, The Johns Hopkins University, the Korean Scientist Group, Los Alamos National Laboratory, the Max-Planck-Institute for Astronomy (MPIA), the Max-Planck-Institute for Astrophysics (MPA), New Mexico State University, University of Pittsburgh, University of Portsmouth, Princeton University, the United States Naval Observatory, and the University of Washington.

\bibliography{paper}

\end{document}